\documentclass{aa}  

\usepackage{graphicx}
\usepackage{txfonts}
\usepackage{natbib}
\usepackage{rotating}
\usepackage{color}

\bibpunct{(}{)}{;}{a}{}{,}

\newcommand{\kms}{km\,s$^{-1}$}

\newcommand{\vsini}{$v\sin{i}$}

\begin{document}

\title{The magnetic field of $\zeta$\,Ori\,A\thanks{Based on observations obtained at the T\'elescope Bernard Lyot (USR5026) operated by the Observatoire Midi-Pyr\'en\'ees, Universit\'e de Toulouse (Paul Sabatier), Centre National de la Recherche Scientifique of France.}}

\author{A. Blaz\`ere\inst{1}
\and C. Neiner\inst{1}
\and A. Tkachenko\inst{2}
\and J.-C. Bouret\inst{3}
\and Th. Rivinius\inst{4}
\and the MiMeS collaboration
}

\institute{LESIA, Observatoire de Paris, PSL Research University, CNRS, Sorbonne Universit\'es, UPMC Univ. Paris 06, Univ. Paris Diderot, Sorbonne Paris Cit\'e, 5 place Jules Janssen, 92195 Meudon, France\\
\email{aurore.blazere@obspm.fr}
\and Instituut voor Sterrenkunde, KU Leuven, Celestijnenlaan 200D, B-3001 Leuven, Belgium
\and Aix-Marseille University, CNRS, LAM (Laboratoire d'Astrophysique de Marseille), UMR 7326, 13388 Marseille, France
\and ESO - European Organisation for Astron. Research in the Southern Hemisphere, Casilla 19001, Santiago, Chile
}

\date{Received ...; accepted ...}

\abstract
{$\zeta$\,Ori\,A is a hot star claimed to host a weak magnetic field, but no clear magnetic detection was obtained so far. In addition, it was recently shown to be a binary system composed of a O9.5I supergiant and a B1IV star.}
{We aim at verifying the presence of a magnetic field in $\zeta$\,Ori\,A, identifying to which of the two binary components it belongs (or whether both stars are magnetic), and characterizing the field.}
{Very high signal-to-noise spectropolarimetric data were obtained with Narval at the Bernard Lyot Telescope (TBL) in France. Archival HEROS, FEROS and UVES spectroscopic data were also used. The data were first disentangled to separate the two components. We then analyzed them with the Least-Squares Deconvolution (LSD) technique to extract the magnetic information. }
{We confirm that $\zeta$\,Ori\,A is magnetic. We find that the supergiant component $\zeta$\,Ori\,Aa is the magnetic component: Zeeman signatures are observed and rotational modulation of the longitudinal magnetic field is clearly detected with a period of 6.829 d. This is the only magnetic O supergiant known as of today. With an oblique dipole field model of the Stokes V profiles, we show that the polar field strength is $\sim$140 G. Because the magnetic field is weak and the stellar wind is strong, $\zeta$\,Ori\,Aa does not host a centrifugally supported magnetosphere. It may host a dynamical magnetosphere. Its companion $\zeta$\,Ori\,Ab does not show any magnetic signature, with an upper limit on the undetected field of $\sim$ 300 G.} 
{}


\maketitle
%

\section{Introduction}

Magnetic fields play a significant role in the evolution of hot massive stars.
However, the basic properties of the magnetic fields of massive stars are poorly
known. About 7\% of the massive stars are found to be magnetic at a level that is detectable
with current instrumentation \citep{wade14b}. In particular, only 11 magnetic
O stars are known. Detecting magnetic field in O stars is particularly
challenging  because they only have few, often broad, lines from which to
measure the field. There is therefore a deficit in the knowledge of the basic
magnetic properties of O stars.

We here study the O star $\zeta$\,Ori\,A. A magnetic field seems to
have been detected in this star by \cite{bouret08}. Their detailed spectroscopic
study of the stellar parameters led to the determination of an effective
temperature of $T_{\rm eff}$ = 29500 $\pm$ 1000 K and log g = 3.25 $\pm$ 0.10
with solar abundances. This makes $\zeta$\,Ori\,A the only magnetic O
supergiant. Moreover, \cite{bouret08} found a magnetic field of 61 $\pm$ 10 G,
which makes it the weakest ever reported field in a hot massive star (typically
ten times weaker than those detected in other magnetic massive stars). They found
a rotational period of $\sim$7 days from the temporal variability of spectral
lines and the modulation of the Zeeman signatures. To derive the magnetic
properties, they used six lines that are not or only weakly affected by the wind.
The rotation period they obtained is compatible with their measured
$v\sin{i}$=100 km~s$^{-1}$.

In addition, the measurement of the magnetic field provided by \cite{bouret08}
allows characterizing the magnetosphere of $\zeta$\,Ori\,A and locating it
in the magnetic confinement-rotation diagram \citep{Petit13}: $\zeta$\,Ori\,A is
the only known magnetic massive star with a confinement parameter below 1, that is,
without a magnetosphere.

For all these reasons, the study of the magnetic field of $\zeta$\,Ori\,A is of
the highest importance. Each massive star that is detected to be
magnetic moves us closer to understanding the stellar magnetism of hot stars. Studying this unique magnetic massive
supergiant is also of particular relevance for our understanding of the
evolution of the magnetic field in hot stars. 

$\zeta$\,Ori\,A has a known B0III companion, $\zeta$\,Ori\,B. In addition,
\cite{hummel13} found that $\zeta$\,Ori\,A consists of two companion stars
located at 40 mas of each other, orbiting with a period of 2687.3$\pm$7.0 days.
To determine a dynamical mass of the components, \cite{hummel13} analyzed
archival spectra to measure the radial velocity variations. The conclusions
reached are presented below. The primary $\zeta$\,Ori\,Aa is a O9.5I
supergiant star, whose radius is estimated to 20.0$\pm$3.2 $R_{\odot}$ and whose
mass is estimated to 33$\pm$10 $M_{\odot}$. The secondary $\zeta$\,Ori\,Ab is a
B1IV with an estimated radius of 7.3$\pm$1.0 $R_{\odot}$ and an estimated mass
of 14$\pm$3 $R_{\odot}$. Moreover, $\zeta$\,Ori\,A is situated at a distance of
387 pc. Initial estimates of the elements of the apparent orbit were obtained by
\cite{hummel13} using the Thiele-Innes method. The estimation provided a value
of the periastron epoch of JD 2452734.2$\pm$9.0 with a longitude of
24.2$\pm$1.2$^{\circ}$. The eccentricity is estimated to be 0.338$\pm$0.004.

\cite{bouret08} considered $\zeta$\,Ori\,A as a single star of 40
$M_{\odot}$ with a radius equal to 25 $R_{\odot}$, seen from Earth at an
inclination angle of 40$^{\circ}$. Taking into account that the star is a binary
could strongly modify the magnetic field value derived for only one of the
binary components. In their analysis, the magnetic signature was
normalized by the full intensity of the lines from both components, and if only
one of the two stars is magnetic, the field was thus underestimated. Moreover,
the position in the magnetic confinement-rotation diagram will be modified as a result of the new magnetic strength value, but also as a consequence of the new stellar parameters. 

Based on new spectropolarimetric observations of $\zeta$\,Ori\,A
and archival spectra presented in Sect.~\ref{sect_obs}, we here seek to confirm that
$\zeta$\,Ori\,A is a magnetic star (Sect.~\ref{sect_mag}). We determine with several techniques, including  by disentangling the composite spectrum
(Sect.~\ref{sect_disentangling}) whether the
magnetic field is hosted by the primary or the secondary star of $\zeta$\,Ori\,A. We then determine the field strength of the
magnetic component (Sect.~\ref{sect_field}) and quantify the non-detection of a
field in the companion (Sect.~\ref{sect_Ab}). In addition, we investigate the
rotational modulation of the magnetic field, its configuration (Sect.~\ref{sect_config}), and the possible
presence of a magnetosphere (Sect.~\ref{sect_magnetosphere}). Finally, we discuss our
results and draw conclusions in Sect.~\ref{sect_discussion}.

\section{Observations}\label{sect_obs}

\subsection{Narval spectropolarimetric observations}

\begin{table}[t]
\centering
\caption{Journal of Narval observations. The columns list the date and the heliocentric Julian date (HJD) for the middle of observation, the number of sequences, and the exposure time per individual subexposure, the signal-to-noise ratio in the I profiles, and the orbital phase.}
\label{obs}
\begin{tabular}{llllll}
\hline
\# & Date & mid-HJD & $T_{exp}$ (s) & S/N & $\phi_{\rm orb}$\\
\hline
\hline
 1 & 17oct07 & 2454391.559 & 48 $\times$ 4 $\times$ 20 & 4750 & 0.617 \\
 2 & 18oct07 & 2454392.719 &  8 $\times$ 4 $\times$ 40 & 2220 & 0.617 \\
 3 & 19oct07 & 2454393.570 & 44 $\times$ 4 $\times$ 40 & 6940 & 0.617 \\
 4 & 20oct07 & 2454394.491 & 48 $\times$ 4 $\times$ 40 & 6860 & 0.618 \\
 5 & 21oct07 & 2454395.518 & 48 $\times$ 4 $\times$ 40 & 7070 & 0.618\\
 6 & 23oct07 & 2454397.496 & 48 $\times$ 4 $\times$ 40 & 7180 & 0.619\\
 7 & 24oct07 & 2454398.526 & 48 $\times$ 4 $\times$ 40 & 7270 & 0.619\\
 8 & 22oct08 & 2454762.644 & 40 $\times$ 4 $\times$ 50 & 6660 & 0.755\\
 9 & 23oct08 & 2454763.645 & 38 $\times$ 4 $\times$ 50 & 5530 & 0.755\\
10 & 24oct08 & 2454764.654 & 36 $\times$ 4 $\times$ 50 & 6790 & 0.756\\
11 & 25oct08 & 2454765.639 & 37 $\times$ 4 $\times$ 50 & 6140 & 0.756\\
12 & 26oct08 & 2454766.635 & 38 $\times$ 4 $\times$ 50 & 6420 & 0.756\\
13 & 04oct11 & 2455839.688 & 12 $\times$ 4 $\times$ 90 & 5810 & 0.156\\
14 & 05oct11 & 2455840.670 & 12 $\times$ 4 $\times$ 90 & 5790 & 0.156\\
15 & 10oct11 & 2455845.608 & 12 $\times$ 4 $\times$ 90 & 2040 & 0.158\\
16 & 11oct11 & 2455846.632 & 12 $\times$ 4 $\times$ 90 & 3450 & 0.158\\
17 & 30oct11 & 2455865.712 & 12 $\times$ 4 $\times$ 90 & 5610 & 0.165\\
18 & 07nov11 & 2455873.557 &  5 $\times$ 4 $\times$ 90 & 2700 & 0.168\\
19 & 11nov11 & 2455877.626 & 12 $\times$ 4 $\times$ 90 & 4860 & 0.170\\
20 & 12nov11 & 2455878.565 & 12 $\times$ 4 $\times$ 90 & 4830 & 0.170\\
21 & 24nov11 & 2455890.673 & 12 $\times$ 4 $\times$ 90 & 4180 & 0.175\\
22 & 25nov11 & 2455891.660 & 12 $\times$ 4 $\times$ 90 & 4900 & 0.175\\
23 & 26nov11 & 2455892.502 & 12 $\times$ 4 $\times$ 90 & 4490 & 0.175\\
24 & 29nov11 & 2455895.667 & 12 $\times$ 4 $\times$ 90 & 5400 & 0.176\\
25 & 30nov11 & 2455896.600 &  6 $\times$ 4 $\times$ 90 & 2030 & 0.177\\
26 & 14dec11 & 2455910.477 & 12 $\times$ 4 $\times$ 90 & 1360 & 0.182\\
27 & 08jan12 & 2455935.555 & 12 $\times$ 4 $\times$ 90 & 5630 & 0.191\\
28 & 13jan12 & 2455940.536 & 12 $\times$ 4 $\times$ 90 & 5060 & 0.193\\
29 & 14jan12 & 2455941.539 & 12 $\times$ 4 $\times$ 90 & 5350 & 0.193\\
30 & 15jan12 & 2455942.475 & 12 $\times$ 4 $\times$ 90 & 4680 & 0.194\\
31 & 16jan12 & 2455943.367 & 12 $\times$ 4 $\times$ 90 & 4520 & 0.194\\
32 & 25jan12 & 2455952.529 & 12 $\times$ 4 $\times$ 90 & 5120 & 0.198\\
33 & 26jan12 & 2455953.431 & 8 $\times$ 4 $\times$ 90 & 3200 & 0.198\\
34 & 08feb12 & 2455966.472 & 12 $\times$ 4 $\times$ 90 & 3900 & 0.203\\
35 & 09feb12 & 2455967.402 & 11 $\times$ 4 $\times$ 120 & 4340 & 0.203\\
36 & 10feb12 & 2455968.343 & 12 $\times$ 4 $\times$ 90 & 2198 & 0.203\\
\hline
\end{tabular}
\end{table}

Spectropolarimetric data of $\zeta$\,Ori\,A were collected with Narval in the
frame of the project Magnetism in Massive Stars (MimeS)  \citep[see
e.g.][]{neiner2011}. This is the same instrument with which the magnetic field of
$\zeta$\,Ori\,A was discovered by \cite{bouret08}. Narval is a
spectropolarimeter installed on the two-meter Bernard Lyot Telescope (TBL) at the
summit of the Pic du Midi in the French Pyr\'en\'ees. This fibre-fed
spectropolarimeter (designed and optimized to detect stellar magnetic
fields through the polarization they generate) provides complete coverage of the
optical spectrum from 3700 to 10500 $\AA$ on 40 echelle orders with a spectral
resolution of $\sim$65000. Considering the size of the fiber, the light from
$\zeta$\,Ori\,B was not recorded in the spectra, but light from both
components of $\zeta$\,Ori\,A was collected.

$\zeta$\,Ori\,A was first observed in October 2007 during 7 nights (PI: J.-C.
Bouret) and these data were used in \cite{bouret08}. Then, this star was observed
again in October 2008 during 5 nights (PI: J.-C. Bouret) and between October
2011 and February 2012 during 24 nights by the MiMeS collaboration (PI: C.
Neiner). This provides a total number of 36 nights of observations. The
observations were taken in circular polarimetric mode, that is, measuring Stokes V.
Each measurement was divided into four subexposures with a different polarimeter
configuration.

Since $\zeta$\,Ori\,A is very bright (V=1.77), only a very short exposure time
could be used to avoid saturation. To increase the total signal-to-noise ratio
(S/N), we thus obtained a number of successive measurements each night, which
were co-added. The exposure time of each subexposure of each measurement varies
between 20 and 120 s, and the total integration time for a night varies between
1280 and 7680 s (see Table~\ref{obs}).

Data were reduced at the telescope using the Libre-Esprit reduction package
\citep{donati97}. We then normalized each of the 40 echelle orders of each of
the 756 spectra with the continuum task of IRAF\footnote{IRAF is distributed by
the National Optical Astronomy Observatories, which are operated by the
Association of Universities for Research in Astronomy, Inc., under cooperative
agreement with the National Science Foundation.}. Finally, we co-added all the
spectra obtained within each night to improve the S/N, which varies between 1360
and 7270 in the intensity spectra (see Table~\ref{obs}). We therefore obtained
36 nightly averaged measurements.

\subsection{Archival spectroscopic observations}\label{sect_spectro}

\begin{figure*}
\centering
\resizebox{\hsize}{!}{\includegraphics[clip, trim=0cm 5cm 3cm 0cm]{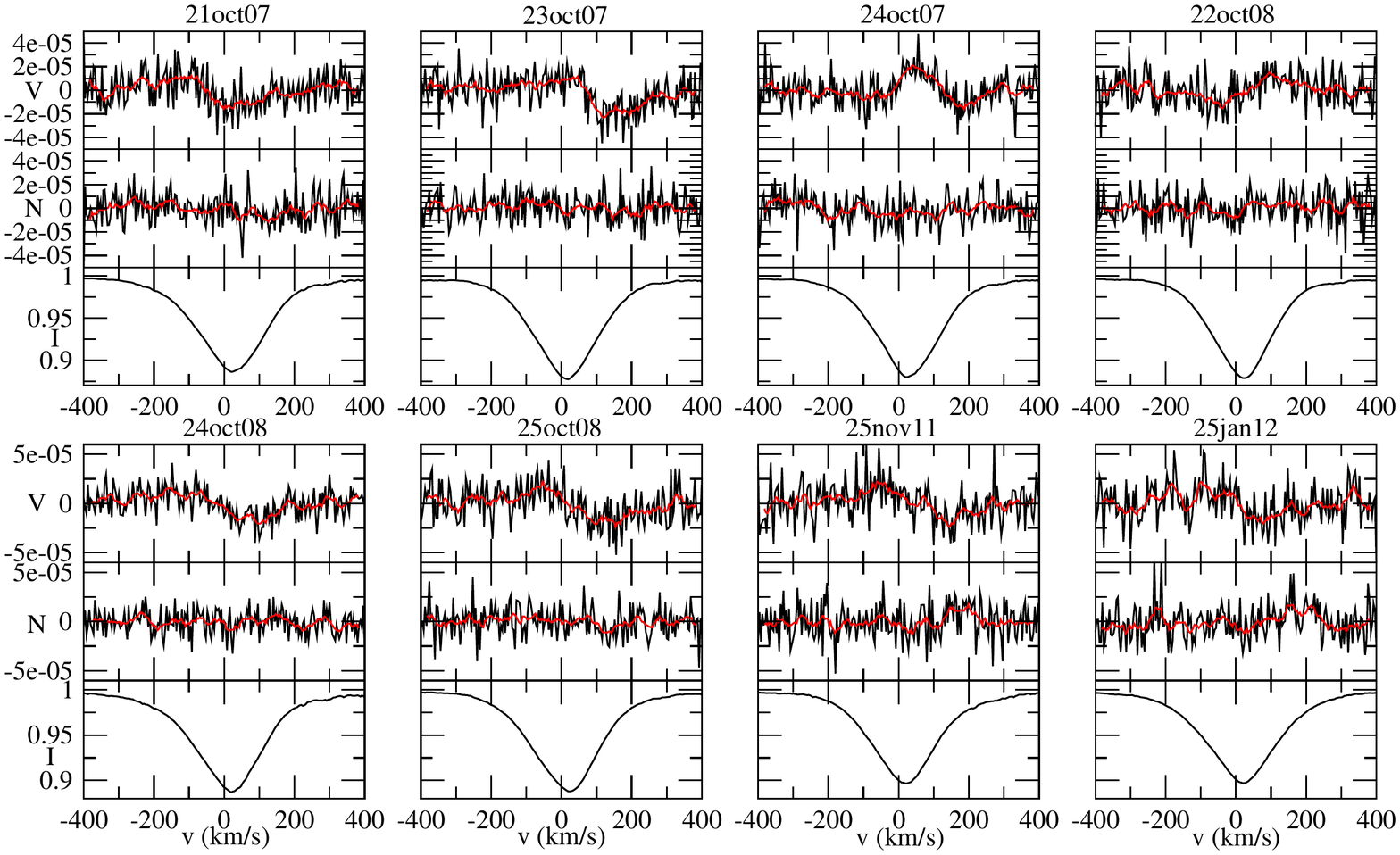}}
\caption{LSD Stokes I (bottom), Stokes V (top), and null N (middle) profiles, normalized to Ic, for 8 selected nights. The red line is a smoothed profile.}
\label{v_n1}
\end{figure*}

In addition to the spectropolarimetric data, we used archival spectroscopic data
of $\zeta$\,Ori\,A taken with various echelle spectrographs.

In 1995, 1997 and 1999, spectra were obtained with the HEROS instrument,
installed at the ESO Dutch 0.9 m telescope at the La Silla Observatory. The
spectral resolution of HEROS is 20000, with a spectral domain from about 350 to
870 nm. In addition, in 2006, 2007 and 2009, data were taken with the FEROS
spectrograph installed at the ESO 2.2 m at the La Silla observatory. The
spectral resolution of FEROS is about 48000 and the spectral domain ranges from
about 370 to 900 nm. Finally, in 2010, spectra were taken with the UVES
spectrograph \citep{dekker00} installed at the VLT at the Paranal Observatory. Its spectral domain ranges from about 300 to 1100 nm with a spectral resolution of 80000 and 110000 in the blue and red domains respectively.

We co-added spectra collected for each year to improve the final S/N. We
therefore have seven spectra for seven different years, with a S/N of between about 100 and
2000 (see Table~\ref{spectro}).

\begin{table}[t]
\centering
\caption{Journal of archival spectroscopic observations of $\zeta$\,Ori\,A obtained with HEROS, FEROS and UVES, showing the date and heliocentric Julian date, instrument used, exposure time, signal-to-noise ratio, and orbital phase.}
\label{spectro}
\begin{tabular}{llllll}
\hline
Date & JD & Instrument & T$_{\rm exp}$ & S/N & $\phi_{\rm orb}$\\
\hline
\hline
 1995 & 2449776.024 & HEROS & 57$\times$1200 & 1200 & 0.90\\
 1997 & 2450454.379 & HEROS & 16$\times$1200 & 1000 & 0.15\\
 1999 & 2451147.333 & HEROS & 64$\times$1200 & 1200 & 0.41\\
 2006 & 2453738.159 & FEROS & 60 &  100 & 0.37\\
 2007 & 2454501.018 & FEROS & 2$\times$20 & 250  & 0.66\\
 2009 & 2454953.970 & FEROS & 5$\times$10 & 200 & 0.84\\
 2010 & 2455435.373 & UVES  & 36$\times$2 & 2000 & 0.01\\
\hline
\end{tabular}
\end{table}

\section{Checking for the presence of a magnetic field}\label{sect_mag}

The magnetic field of $\zeta$\,Ori\,A claimed by \cite{bouret08} has not been
confirmed by independent observations so far and one of the goals of this study
is to confirm or disprove its existence using additional observations.

To test whether $\zeta$\,Ori\,A is magnetic, we applied the Least-Squares
Deconvolution (LSD) technique \citep{donati97}. We first created a line mask for
$\zeta$\,Ori\,A. We started from a list of lines extracted from VALD
\citep{piskunov95, kupka99} for an O star with $T_{\rm eff}$=30000 K and log
g=3.25, with their Land\'e factors and theoretical line depths. We then cleaned
this line list by removing the hydrogen lines, the lines that are blended with
hydrogen lines, and those that are not visible in the spectra. We also
added some lines visible in the spectra that were not in the original O-star
mask. Altogether, we obtained a mask of 210 lines. We then adjusted the depth of
these 210 lines in the mask to fit the observed line depths.

Using the final line mask, we extracted LSD Stokes I and V profiles for each
night. We also extracted null (N) polarization profiles to check for spurious
signatures. The LSD I, Stokes V and the null N profiles are shown in
Fig.~\ref{v_n1} for 8 of 36 nights. A plot of all profiles is available
online (Fig.~\ref{lsd3}). Zeeman signatures are clearly seen for these 8  nights
and some others as well, but are not systematically observed for all nights. 
We calculated the false alarm probability (FAP) by comparing the signal inside the lines with no signal \citep{donati97}. If FAP $<$ 0.001\%, the magnetic detection
is definite; if it is 0.001\% $<$ FAP $<$ 0.1\% the detection is marginal, otherwise
there is no detection. Table~\ref{bl} indicates whether a definite detection
(DD), marginal detection (MD) or no detection (ND) was obtained for each of the
night. DD or MD were obtained in 15 out of 36 nights. The existence of Zeeman
signatures confirms that $\zeta$\,Ori\,A hosts a magnetic field, as previously
reported by \cite{bouret08}.

\begin{figure}
\centering
\includegraphics[scale=0.36,clip, trim=1.5cm 1cm 0cm 2.5cm]{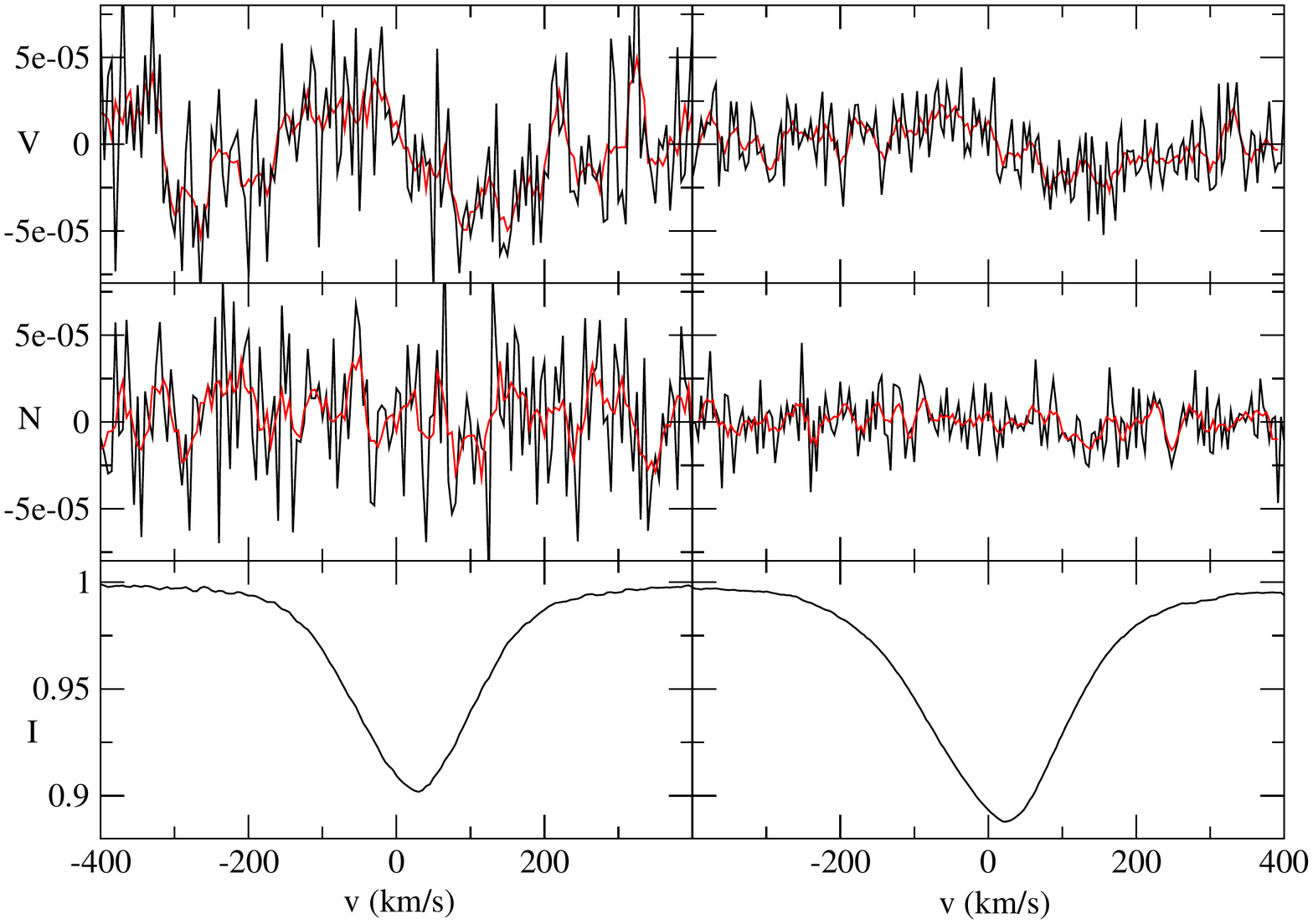}
\caption{LSD Stokes I (bottom), Stokes V (top), and null N (middle) profiles,  normalized to Ic, for the night of 25 October 2008 for the mask of \cite{bouret08} (left) and for our own mask (right). The red line is a smoothed profile.}
\label{compare}
\end{figure}

The previous study of the magnetic field of $\zeta$\,Ori\,A \citep{bouret08} only used a few lines that were not affected by the wind. However, we need to use as many lines as possible to improve the S/N. We therefore checked whethe our results were changed by using lines that might be affected by the wind. We compared the LSD results obtained with the mask used by \cite{bouret08} and our own mask (see Fig.~\ref{compare}). The signatures in Stokes V are similar with both masks and the measurements of the longitudinal magnetic field are consistent (e.g., 44.5$\pm$19.6 G with the mask of \cite{bouret08} and 35.8$\pm$7.2 G with our mask for measurement \# 11, see Fig.~\ref{compare}). However, the S/N is better with our mask (the S/N of Stokes V is 57624) than with the mask of \cite{bouret08} (the S/N of Stokes V is 27296). Therefore, we used all available lines for this study.

However, the line mask used in this first analysis includes lines from both
components of $\zeta$\,Ori\,A. We thus ignored which component of the binary is
magnetic or whether both components are magnetic. To provide an answer to this
question, we must separate the composite spectra.

\section{Separating the two components}\label{sect_disentangling}

\subsection{Identifying the lines of each component}\label{lines}

We first created synthetic spectra of each component. The goal was to identify
which lines come from the primary component, the secondary component, or both.
To this aim, we used TLUSTY \citep{hubeny95}. This program calculates
plane-parallel, horizontally homogeneous stellar atmosphere models in radiative
and hydrostatic equilibrium. One of the most important features of the program
is that it allows for a fully consistent, non-LTE metal line blanketing.
However, TLUSTY does not take winds into account, which can be important in
massive stars, especially in supergiants.

For the primary component $\zeta$\,Ori\,Aa, we computed a model with an
effective temperature $T_{\rm eff}$=29500 K and logg=3.25, corresponding to the
spectral type of the primary as given by \cite{hummel13}. For the secondary, we
computed a model with an effective temperature $T_{\rm eff}$=29000 K et
logg=4.0, again following \cite{hummel13}. We used solar
abundances, for both stars. The emergen spectrum from a given model atmosphere was calculated
with SYNSPEC\footnote{Synspec is a general spectrum synthesis program developed
by Ivan Hubeny and Thierry Lanz:
http://nova.astro.umd.edu/Synspec49/synspec.html}. This program is complemented
by the program ROTINS, which calculates the rotational and instrumental
convolutions for the net spectrum produced by SYNSPEC.

Comparing these two synthetic spectra to the observed spectra of
$\zeta$\,Ori\,A, we identified which lines belong only to $\zeta$\,Ori\,Aa, only to
$\zeta$\,Ori\,Ab, and which are a blend of the lines of both components. 
If one observed line only existed in one synthetic spectrum, we considered that this line is only emitted from one component of the binary. If it existed in both synthetic spectra, we considered this line to be a blend of both components.
We then
created line lists containing lines from the three categories (only Aa, only Ab, or both).

In addition, we gathered archival spectra of $\zeta$\,Ori\,A taken with the
spectrographs FEROS, HEROS and UVES (see Sect.~\ref{sect_spectro}). While these
spectra do not include polarimetric information, they  cover the orbital period
much better than our Narval data (see Table~\ref{spectro} and Fig.~\ref{phase}).
In particular, some spectra were obtained close to the maximum or minimum of the radial velocity
(RV) curve.

We compared the spectrum taken close to the maximum and minimum RV, because the
line shift is maximum between these two spectra. We  arbitrarily decided to use
the spectra taken close to the maximum as reference. Depending on the shift, we
determined the origin of the lines. If the lines of the spectrum taken at
minimum RV are shifted to the blue (respectively red) side compared to the
spectrum at maximum RV, the line comes from the primary Aa (respectively
secondary Ab) component. When lines from the two components are blended, the
core of the lines are shifted to the red side and the wings to the blue side. 

The identification of lines made this way resulted in similar line lists as
those obtained by comparing the observed spectra with synthetic ones.

We then ran LSD again on the observed Narval spectra, once with the mask containing the
157 lines only belonging to $\zeta$\,Ori\,Aa and once with the mask only containing the
67 lines from $\zeta$\,Ori\,Ab. We observe magnetic signatures in the LSD V
profiles of $\zeta$\,Ori\,Aa that are similar to those obtained in the original LSD
analysis presented in Sect.~\ref{sect_mag}. In the contrast, we do not observe
magnetic signatures in the LSD Stokes V profiles of $\zeta$\,Ori\,Ab. We
conclude that $\zeta$\,Ori\,Aa is magnetic and $\zeta$\,Ori\,Ab is not.

However, the LSD profiles of $\zeta$\, Ori\,Aa obtained this way are very noisy,
because of the low number of lines in the mask, and they cannot be used to precisely estimate the
longitudinal magnetic field strength. To go further, it is necessary to disentangle the spectra, so that more lines can be used.

\subsection{Spectral disentangling of Narval data}\label{dis_narval}

We first attempted to use the Fourier-based formulation of the spectral
disentangling (hereafter, {\sc spd}) method \citep{hadrava95} as implemented in
the {\sc FDBinary} code \citep{ilijic04} to simultaneously determine the orbital
elements and the individual spectra of the two components Aa and Ab of the
$\zeta$\,Ori\,A binary system. The Fourier-based {\sc spd} method is superior to
the original formulation presented by \citet{simon94} that is applied in the wavelength
domain in that it is less time-consuming. In particular, this increases the
technique's efficiency when it is applied to long time-series of high-resolution
spectroscopic data.

One of the pre-conditions for the {\sc spd} method to work efficiently is a
homogeneous phase coverage of the orbital cycle with the data. In particular,
covering the regions of maximum/minimum radial velocity (RV) separation of the
two stars is essential, because these phases provide key information about the RV
semi-amplitudes of both stellar components.

\begin{figure}
\includegraphics[scale=0.5, trim=3cm 13cm 0cm 4cm,clip]{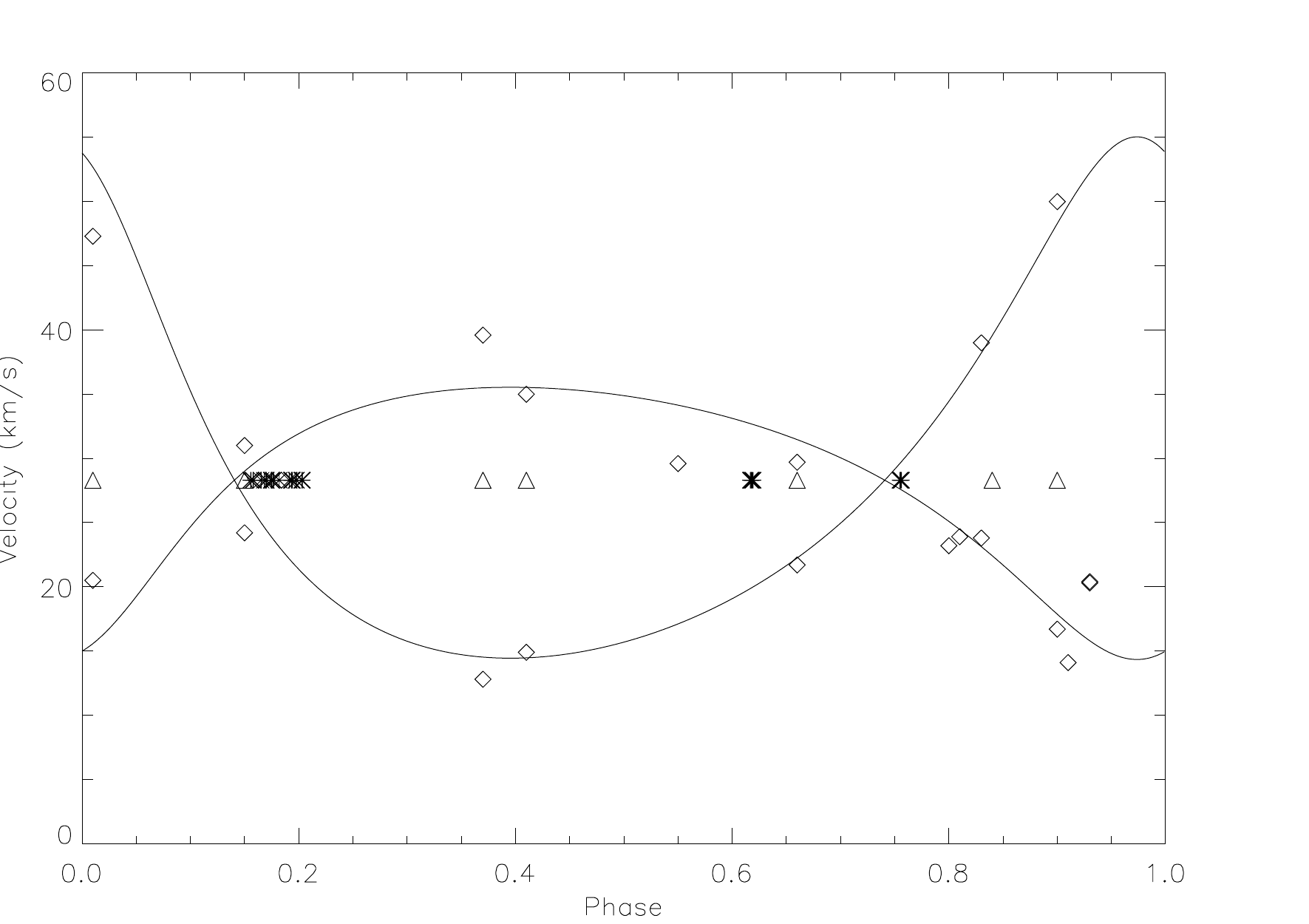}
\caption{Orbital phase distribution of the spectra of $\zeta$\,Ori\,A. The
diamonds indicate the radial velocity measured by \cite{hummel13}. The lines
correspond to the best fit of the radial velocity of each component. The crosses
correspond to our Narval observations and the triangles to the archival spectroscopic data. Phase zero corresponds to the time of
periastron passage (T$_0$ = 2452734.2~HJD) as reported by \citet{hummel13}.}
\label{phase}
\end{figure}

Figure~\ref{phase} illustrates the phase distribution of our Narval spectra
according to the orbital period of 2687.3~days reported by \citet{hummel13}.
Obviously, the spectra provide very poor phase coverage; no measurements exist
at phases $\sim$0.0 and 0.4, corresponding to a maximum RV separation of
the components \citep[see Fig.~5 in][]{hummel13}. This prevents determining
of accurate orbital elements from our Narval spectra.

Our attempt to use the orbital solution obtained by \citet{hummel13} to separate
the spectra of the individual components also failed: although all regions in
which we disentangled the spectra indicate the presence of lines from the secondary
in the composite spectra, the separated spectra themselves are unreliable.

\subsection{Disentangling using the archival spectroscopic data}

Since the {\sc spd} method failed in disentangling the Narval spectra because of the poor
phase coverage, we again used the spectroscopic archival data obtained with
FEROS, HEROS, and UVES. The orbital coverage of these spectra is much better than
the one of the Narval data. We have seven spectra taken at different orbital phases,
including phases of maximum RV separation of the components (see
Table~\ref{phase}). We first normalized the spectra with IRAF. We used the orbital parameters given by \cite{hummel13} for the disentangling.

\begin{figure}
\includegraphics[scale=0.39,trim=2cm 1.0cm 2cm 2.5cm,clip]{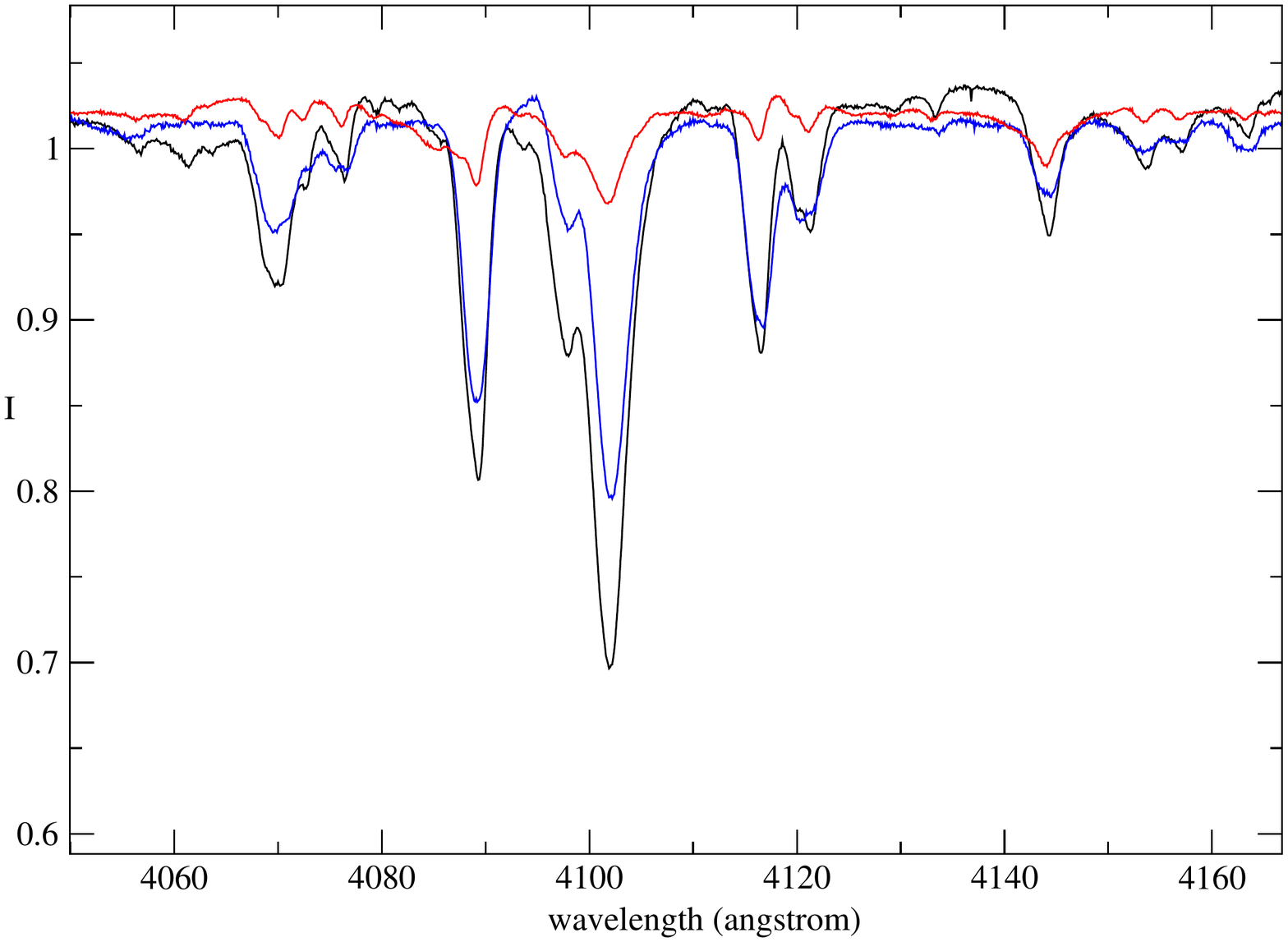}
\caption{Small part of the spectrum of $\zeta$\,Ori\,A showing the composite observed spectrum (black), the spectrum of $\zeta$\,Ori\,Aa (blue) and that of $\zeta$\,Ori\,Ab (red).}
\label{disentangle}
\end{figure}

The coverage of these spectra enables the disentangling using {\sc FDBinary}. As an
illustration of the results, a small part of the disentangled spectra is shown
in Fig.~\ref{disentangle}. The results confirms the origin
of the lines that were determined in Sect.\ref{lines}, and also the spectral types of
the components given by \cite{hummel13}.

\section{Measuring the longitudinal magnetic field of
$\zeta$\,Ori\,Aa}\label{sect_field}

Following these results, we assume that $\zeta$\,Ori\,Ab is not
magnetic and that the Stokes V signal only comes from $\zeta$\,Ori\,Aa.
Therefore, we ran the LSD technique on the Narval spectra with a mask
containing all lines emitted from $\zeta$\,Ori\,Aa, even those that are blended with the
ones of $\zeta$\,Ori\,Ab, to obtain the LSD Stokes V profile of
$\zeta$\,Ori\,Aa. The contribution of $\zeta$\,Ori\,Ab to this Stokes V signal
will be null, as the magnetic signal is only provided by $\zeta$\,Ori\,Aa.

However, we were unable to disentangle the Narval data (see
Sect.~\ref{dis_narval}), therefore the LSD Stokes I spectra of $\zeta$\,Ori\,Aa
could not be computed in the same way as the LSD Stokes V spectra. As a
consequence, we attempted to compute the LSD Stokes I profiles in several ways.

\subsection{Using the Narval data and correcting for the companion}

For the I profiles, we first proceeded in the following way: we computed the LSD
Stokes I profiles with different masks that only contained the lines of $\zeta$\,Ori\,Aa, only the lines of $\zeta$\,Ori\,Ab, and the only blended lines. We
subtracted the LSD Stokes I profiles obtained for the lines of $\zeta$\,Ori\,Ab
alone from the LSD Stokes I profiles obtained for blended lines to remove the
contribution from the Ab component. We then averaged the LSD Stokes I profiles
obtained this way and the one obtained for the lines of $\zeta$\,Ori\,Aa alone.
In this way the same list of lines (those of Aa alone and the blended ones) are
used in the final LSD Stokes I profiles as in the LSD Stokes V profiles
calculated above. 

This allowed us to use more lines than in Sect.~\ref{lines} (i.e., to include the
blended lines) and to improve the resulting S/N. We obtained magnetic signatures
similar to those derived in Sects.~\ref{sect_mag} and \ref{lines} (see
Fig.~\ref{v_n1}). However, the S/N remained low, and some contribution from the
Ab component is probably still present in the LSD Stokes I profile. Longitudinal
field values extracted from these LSD profiles may thus be unreliable. 

\subsection{Using synthetic intensity profiles}

To improve the LSD I profiles, we attempted to use the synthetic TLUSTY/SYNSPEC
spectra calculated in Sect.~\ref{lines} for $\zeta$\,Ori\,Aa. We ran the LSD
tool on the synthetic spectra to produce the synthetic LSD Stokes I profiles of
$\zeta$\,Ori\,Aa with the same line mask as the one used for the LSD Stokes V
profiles above. 

We then computed the longitudinal magnetic field values from the observed LSD
Stokes V profiles and the synthetic LSD Stokes I profiles.  We calculated the
longitudinal magnetic field B$_l$ for all observations with the center-of-gravity method \citep{rees79},
\[ B_{l}=-2.14\times10^{-11}\frac{\int v V(v)dv}{\lambda_{0}g_{m}c \int (1-I(v))dv}\ G.\]

We obtained  longitudinal magnetic field values between -144 and +112 G with
error bars between 20 and 100 G.

\subsection{Using disentangled spectroscopic data}

Although we were unable to disentangle the Narval data, we obtained
disentangled spectra from the purely spectroscopic archival data. To derive the
longitudinal magnetic field values more accurately, we therefore used the
disentangled spectra obtained from the purely spectroscopic data. 

We ran the LSD technique on the disentangled archival spectra obtained for
$\zeta$\,Ori\,Aa using the same line list as we used for Stokes V. Thus, we
obtained the observed mean intensity profile for $\zeta$\,Ori\,Aa alone. We then computed
the longitudinal magnetic field values from the observed LSD Stokes V profiles
from Narval and the observed LSD Stokes I profiles from the disentangled
spectroscopic spectra. 

\begin{figure}
\centering
\includegraphics[scale=0.37,trim=1.2cm 0.0cm 0cm 2cm,clip]{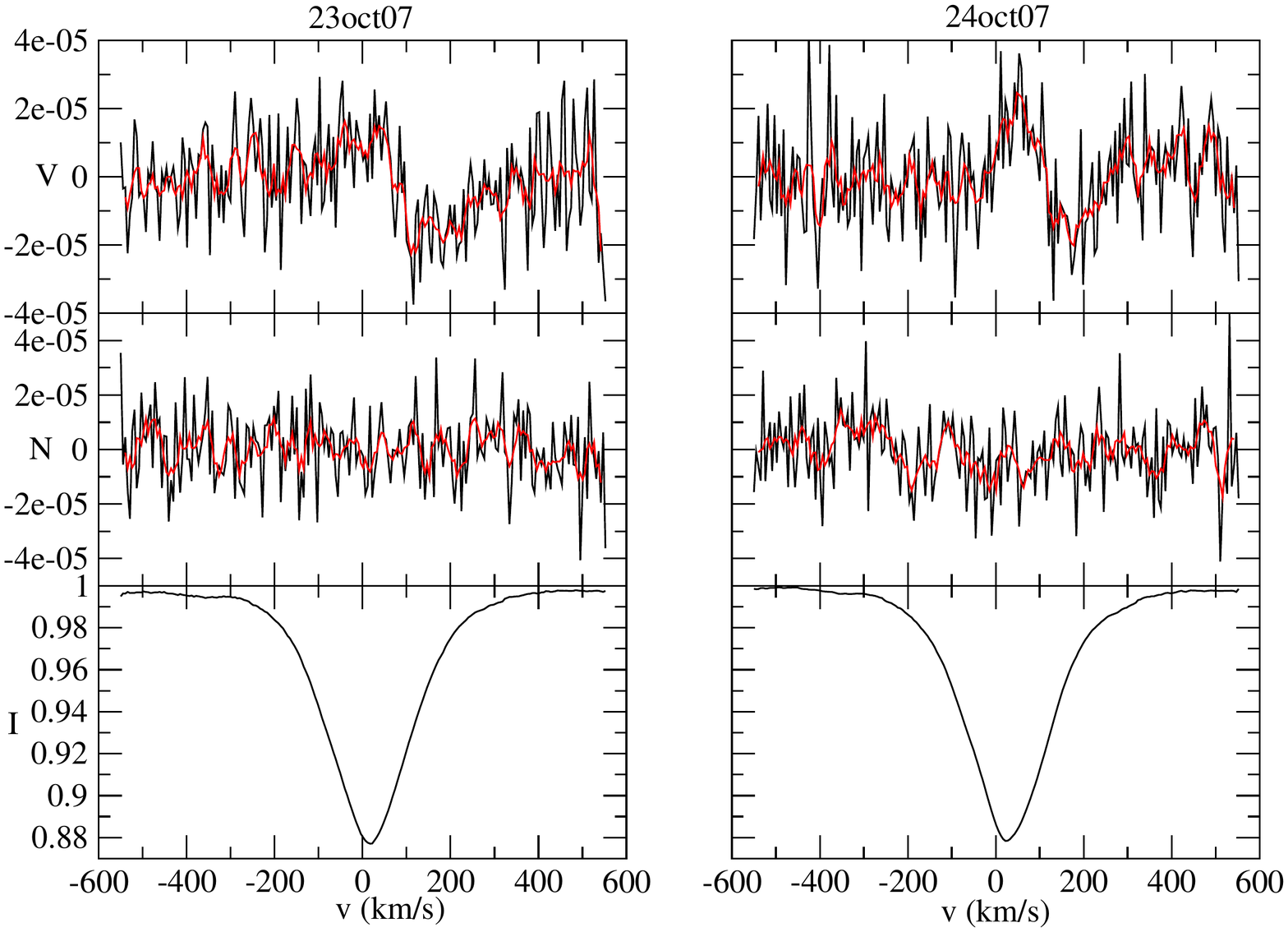}
\includegraphics[scale=0.37,trim=1.2cm 1cm 0cm 2.5cm,clip]{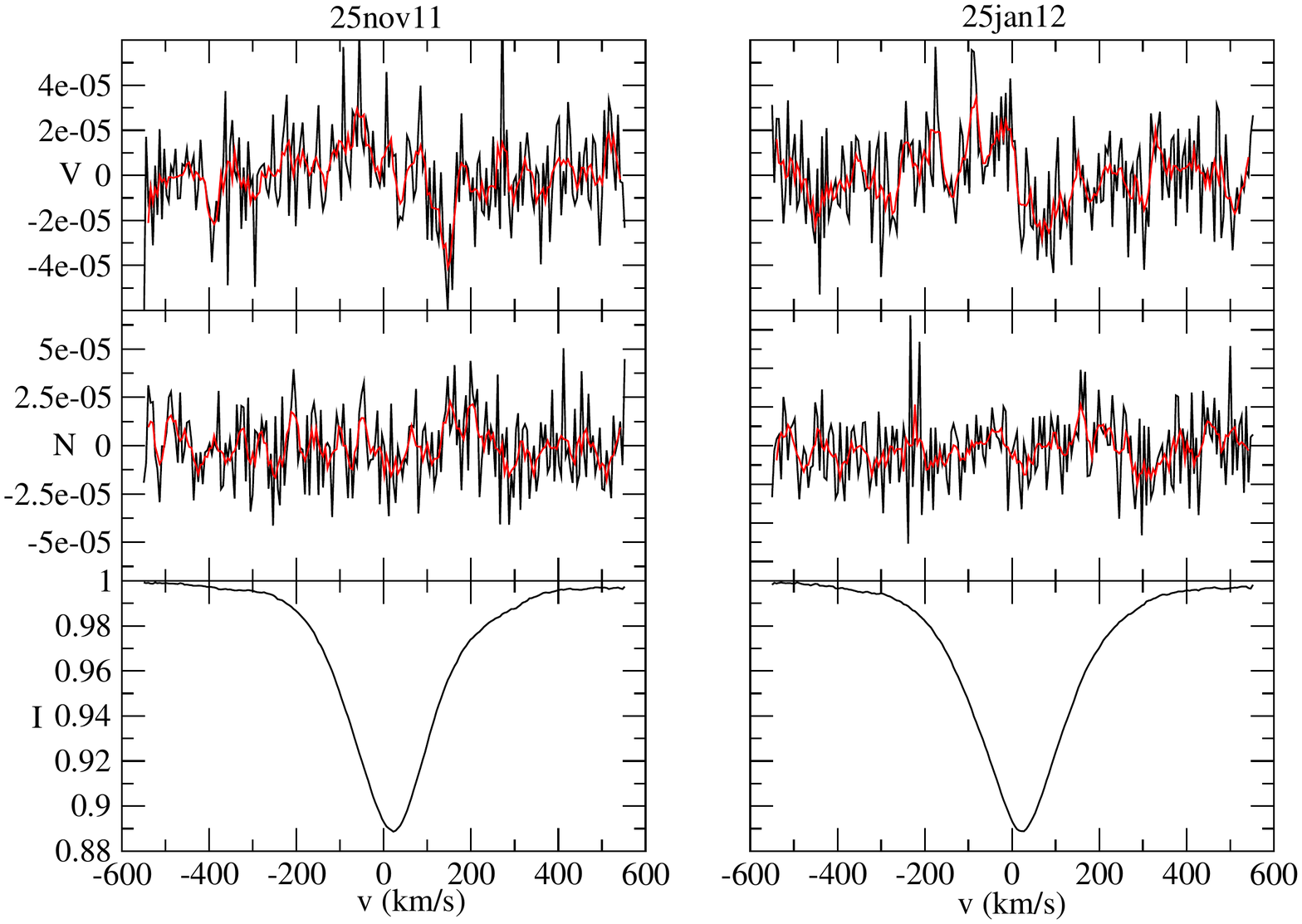}
\caption{Examples of LSD Stokes I profiles (bottom) computed from the disentangled spectroscopic data, Stokes V (top) and null N (middle) profiles, normalized to Ic, from the Narval data for the primary component $\zeta$\,Ori\,Aa for a few  nights of observations. The red line is a smoothed profile.}
\label{lsd_2}
\end{figure}

The shape of the magnetic signatures in LSD Stokes V profiles (Fig.~\ref{lsd_2})
is  similar to the shapes obtained for the combined spectra (Fig.~\ref{v_n1}) and
the various methods presented above. The LSD Stokes I spectra now
better represent the observed $\zeta$\,Ori\,Aa spectrum, however. We therefore adopted these LSD profiles in the remainder of this work. Fifteen of the 36 measurements are DD or MD.

As above, we calculated the longitudinal magnetic field B$_l$ for all
observations with the center-of-gravity method \citep{rees79}. Results are
given in Table~\ref{bl}. The longitudinal field B$_{l}$ varies between about -30
and +50 G, with typical error bars below 10 G. N values are systematically
compatible with 0 within 3$\sigma$N, where $\sigma$N is the error on N, while
B$_{l}$ values are above 3$\sigma$B$_{l}$ in seven instances, where $\sigma$B$_{l}$
is the error on B$_{l}$.

\begin{table}[t]
\centering
\caption{Longitudinal magnetic field of the magnetic primary star $\zeta$\,Ori\,Aa. The columns list the heliocentric Julian dates (HJD) for the middle of observation, the longitudinal magnetic field and its error in gauss, the detection status: definite detection (DD), marginal detection (MD) and no detection (ND), and the "null" polarization and its error in gauss.}
\label{bl}
\begin{tabular}{llllllll}
\hline
\# & mid-HJD & B$_{l}$ & $\sigma$B$_{l}$ & Detect. & N & $\sigma$N\\
\hline
\hline
 1 & 2454391.559 & -5.7  & 7.7  & MD & -3.9 & 7.7 \\
 2 & 2454392.719 & -26.9 & 16.6 & ND & 37.2 & 16.6\\
 3 & 2454393.570 & -9.3  & 5.5  & ND & 3.6 & 5.5\\
 4 & 2454394.491 & -0.3  & 5.4  & ND & 6.3 & 5.4\\
 5 & 2454395.518 & 18.1  & 5.2  & MD & 12.4 & 5.4\\
 6 & 2454397.496 & 25.5  & 5.1  & MD & 3.9 & 5.1\\
 7 & 2454398.526 & 4.7   & 5.1  & MD & -3.7 & 5.1\\
 8 & 2454762.644 & -15.1 & 5.5  & MD & -6.9 & 5.5\\
 9 & 2454763.645 & 12.8  & 6.6  & ND & -5.4 & 6.6\\
10 & 2454764.654 & 28.0  & 5.3  & MD & 3.6 & 5.3 \\
11 & 2454765.639 & 32.8  & 6.3  & DD & 11.1 & 6.3\\
12 & 2454766.635 & 10.9  & 6.4  & MD & 12.7 & 6.4\\
13 & 2455839.688 & -3.9  & 6.5  & ND & 2.9 & 6.5\\
14 & 2455840.670 & 3.9   & 6.5  & ND & -6.7 & 6.6\\
15 & 2455845.608 & 21.3  & 9.4  & ND & 4.8 & 9.4\\
16 & 2455846.632 & -9.7  & 11.0 & ND & -1.8 & 11.0\\
17 & 2455865.712 & 12.4  & 6.5  & ND & -0.9 & 6.6\\
18 & 2455873.557 & 25.2  & 13.5 & MD & -3.6 & 13.5\\
19 & 2455877.626 & 13.6  & 7.5  & ND & -13.5 & 7.5\\
20 & 2455878.565 & 7.7   & 7.5  & ND & 6.8 & 7.5\\
21 & 2455890.673 & 6.1   & 8.7  & ND & 11.0 & 8.7\\
22 & 2455891.660 & 24.0  & 7.4  & MD & -6.7 & 7.4\\
23 & 2455892.502 & 6.0   & 8.1  & ND & -11.4 & 8.1\\
24 & 2455895.667 & 1.3   & 6.9  & ND & -8.0 & 6.9\\
25 & 2455896.600 & -22.6 & 22.3 & ND & 22.3 & 22.5\\
26 & 2455910.477 & 4.1   & 13.8 & ND & 13.6 & 13.8\\
27 & 2455935.555 & -7.2  & 7.0  & MD & 4.2 & 7.0\\
28 & 2455940.536 & 5.9   & 7.2  & ND & -10.1 & 7.2\\
29 & 2455941.539 & -3.1  & 6.8  & MD & 4.5 & 6.8\\
30 & 2455942.475 & -5.3  & 8.0  & ND & -4.4 & 8.0\\
31 & 2455943.367 & 4.8   & 8.4  & ND & 4.3 & 8.4\\
32 & 2455952.529 & 25.2  & 7.3  & MD & -12.5 & 7.3\\
33 & 2455953.431 & 72.3  & 59.1 & DD & -60.4 & 59.1\\
34 & 2455966.472 & 51.0  & 9.5  & MD & -0.3 & 9.5\\
35 & 2455967.402 & 1.7   & 8.5  & ND & 11.3 & 8.4\\
36 & 2455968.343 & 10.7  & 16.7 & ND & -13.7 & 16.6\\
\hline
\end{tabular}
\end{table}

\section{No magnetic field in $\zeta$\,Ori\,Ab}\label{sect_Ab}

\subsection{Longitudinal magnetic field values for $\zeta$\,Ori\,Ab}

\begin{figure}
\centering
\includegraphics[scale=0.37,trim=1cm 0.0cm 0.5cm 2cm,clip]{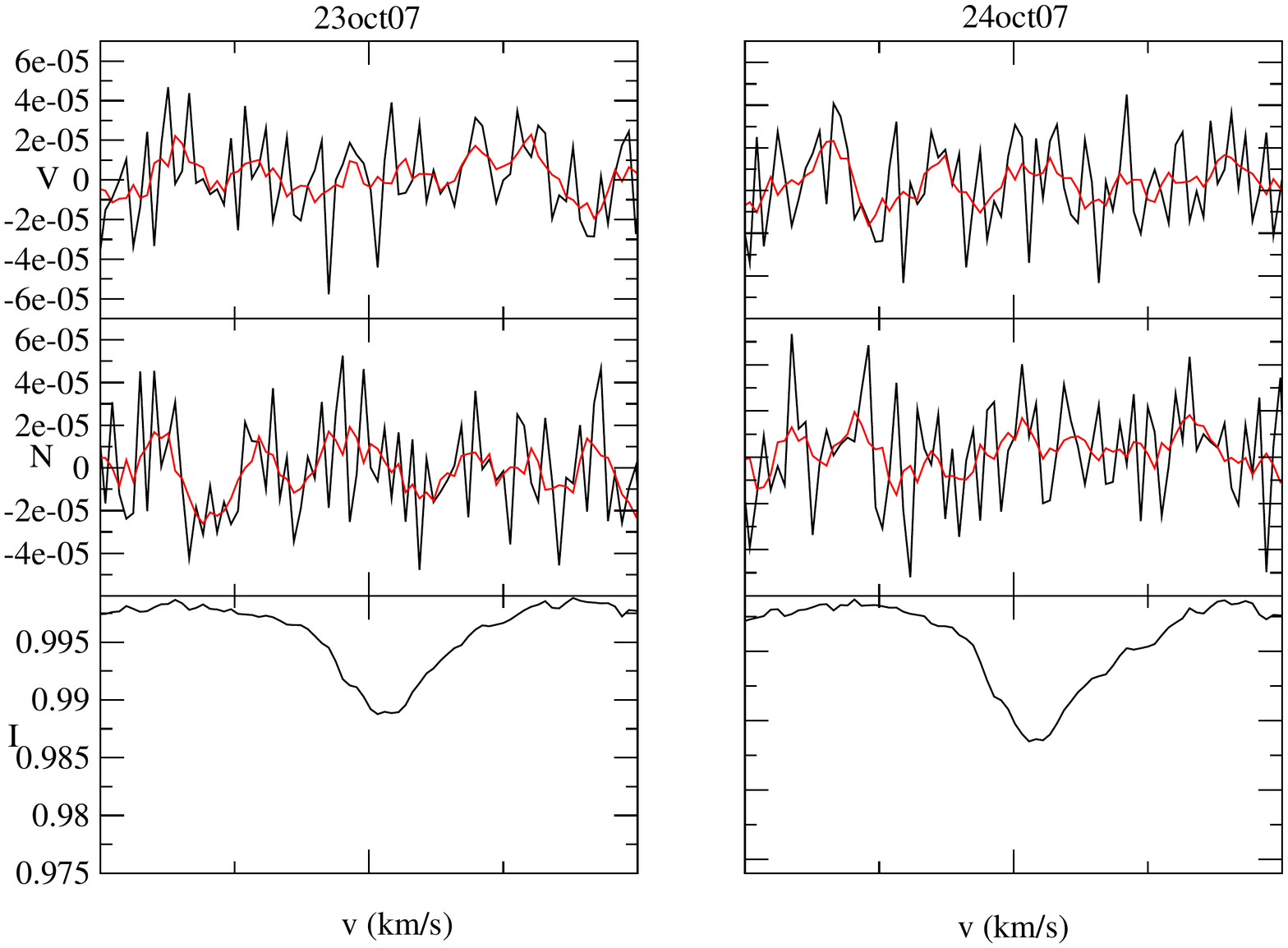}
\includegraphics[scale=0.37,trim=1cm 1cm 0.5cm 2.5cm,clip]{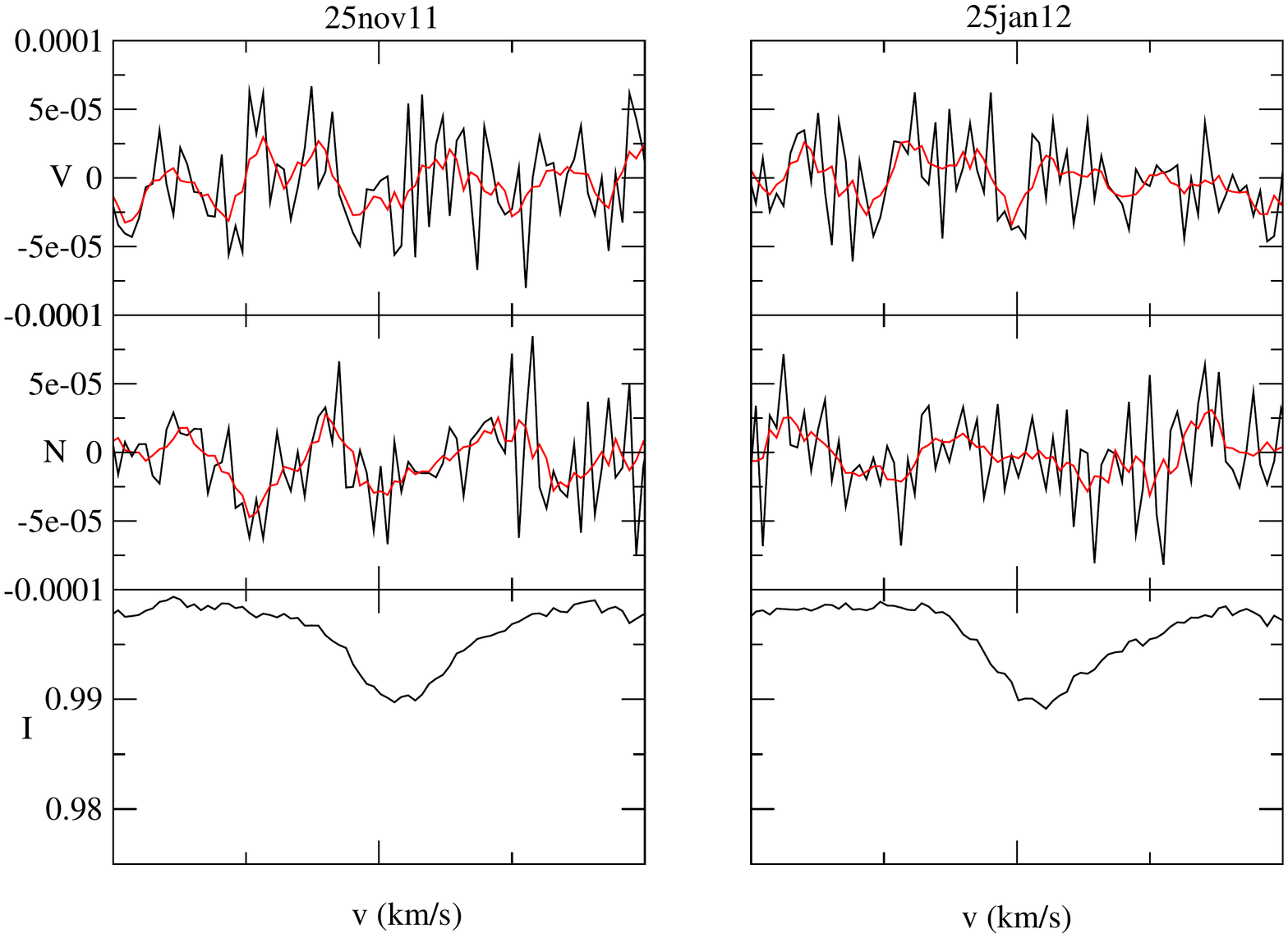}
\caption{Examples of LSD Stokes I (bottom), Stokes V (top), and null N (middle) profiles, normalized to Ic, for the secondary component $\zeta$\,Ori\,Ab for a few nights of observations, computed from Narval data using only the 82 lines belonging to the secondary component. The red line is a smoothed profile.}
\label{lsd_zetaoriab}
\end{figure}

To confirm the non-detection of a magnetic field in $\zeta$\,Ori\,Ab, we ran the
LSD technique with a mask that only contained lines emitted from $\zeta$\,Ori\,Ab,
that is 67 lines. This ensures that the LSD Stokes V profiles are not polluted by
the magnetic field of $\zeta$\,Ori\,Aa. Signatures in the LSD Stokes V profiles
are not detected in any of the profiles (all ND), as shown in Table~\ref{bl_Ab} and in
Fig.~\ref{lsd_zetaoriab} for selected nights when a signal is detected in
$\zeta$\,Ori\,Aa.

Using these LSD profiles and the center-of-gravity method, we calculated the
longitudinal field value, the null polarization, and their error bars for
$\zeta$\,Ori\,Ab. We find that both B$_{l}$ and N are compatible with 0 within
3$\sigma$ for all nights (see Table~\ref{bl_Ab}). However, the error bars on the
longitudinal field values of $\zeta$\,Ori\,Ab are much higher (typically 70 G)
than those for $\zeta$\,Ori\,Aa (typically 10 G), because far fewer lines could
be used to extract the signal for $\zeta$\,Ori\,Ab. 

\begin{table}[t]
\centering
\caption{Longitudinal magnetic field measurements for the secondary $\zeta$\,Ori\,Ab. The columns list the heliocentric Julian dates (HJD) for the middle of observation, the longitudinal magnetic field and its error in gauss, the detection status: no detection (ND) in all cases, and the "null" polarization and its error in gauss.}
\label{bl_Ab}
\begin{tabular}{@{\,}lllll@{\,}l@{\,}ll}
\hline
\# & mid-HJD & B$_{l}$ & $\sigma$B$_{l}$ & Detect. & N & $\sigma$N\\
\hline
\hline
 1 & 2454391.559 & -62.9  & 74.5  & ND & 46.5 & 74.4 \\
 2 & 2454392.719 & -80.4 & 132.7 & ND & 91.9 & 132.8\\
 3 & 2454393.570 & 26.0  & 54.8  & ND & -60.6 & 54.9\\
 4 & 2454394.491 & 31.0  & 56.7  & ND & 24.1 & 56.6\\
 5 & 2454395.518 & 86.7  & 47.6  & ND & 101.4 & 47.7\\
 6 & 2454397.496 & -49.4 & 45.5  & ND & 14.3 & 45.9\\
 7 & 2454398.526 & 14.2  & 46.0  & ND & 44.6 & 46.1\\
 8 & 2454762.644 & -14.2 & 50.8  & ND & 73.0 & 50.6\\
 9 & 2454763.645 & -24.1  & 65.4 & ND & 43.3 & 65.5\\
10 & 2454764.654 & -40.1  & 49.2  & ND & 80.0 & 49.4 \\
11 & 2454765.639 & -32.9 & 73.5  & ND & 47.2 & 73.6\\
12 & 2454766.635 & 7.9  & 60.5  & ND & -26.5 & 60.6\\
13 & 2455839.688 & -62.3  & 53.2  & ND & -30.5 & 53.6\\
14 & 2455840.670 & -8.3  & 58.3  & ND & 3.4 & 58.2\\
15 & 2455845.608 & -19.1  & 92.2  & ND & -21.9 & 91.1\\
16 & 2455846.632 & -126.4  & 91.9 & ND & -102.1 & 91.2\\
17 & 2455865.712 & 35.7  & 63.2  & ND & 173.4 & 63.9\\
18 & 2455873.557 & 141.5  & 112.7 & ND & -224.0 & 113.0\\
19 & 2455877.626 & 55.7  & 60.0  & ND & 2.5 & 60.0\\
20 & 2455878.565 & 34.8   & 68.3  & ND & 75.0 & 68.7\\
21 & 2455890.673 & -105.9   & 106.6  & ND & -154.9 & 107.5\\
22 & 2455891.660 & 8.7  & 82.6  & ND & 86.6 & 82.8\\
23 & 2455892.502 & 77.1   & 84.3  & ND & -63.7 & 84.8\\
24 & 2455895.667 & -27.0   & 75.2  & ND & -63.8 & 75.5\\
25 & 2455896.600 & -33.7 & 289.4 & ND & -113.7 & 296.7\\
26 & 2455910.477 & 171.0   & 183.7 & ND & 129.3 & 183.4\\
27 & 2455935.555 & -82.4  & 70.1  & ND & -58.6 & 69.9\\
28 & 2455940.536 & -12.3   & 63.5  & ND & -55.8 & 63.7\\
29 & 2455941.539 & 20.3  & 67.3  & ND & 116.9 & 67.8\\
30 & 2455942.475 & 31.8  & 70.6  & ND & -97.2 & 71.1\\
31 & 2455943.367 & -6.2   & 78.8  & ND & -27.7 & 79.2\\
32 & 2455952.529 & -72.4  & 71.0  & ND & 1.1 & 71.3\\
33 & 2455953.431 & 173.6  & 478.5 & ND & -152.6 & 480.7\\
34 & 2455966.472 & 52.9  & 79.9  & ND & -32.1 & 79.8\\
35 & 2455967.402 & -57.8   & 74.6  & ND & 62.8 & 74.3\\
36 & 2455968.343 & -152.8  & 186.5 & ND & -92.9 & 187.2\\
\hline
\end{tabular}
\end{table}

\subsection{Upper limit on the non-detected field in $\zeta$\,Ori\,Ab}

The signature of a weak magnetic field might have remained hidden in the noise of
the spectra of the $\zeta$\,Ori\,Ab. To evaluate its maximum strength, 
we first fitted the LSD $I$ profiles computed above for $\zeta$\,Ori\,Ab with a
double Gaussian profile. This fit does not use physical stellar parameters, but
it reproduces the $I$ profiles as well as possible. We then calculated 1000
oblique dipole models of each of the LSD Stokes $V$ profiles for various values
of the polar magnetic field strength  $B_{\rm  pol}$. Each of these models uses
a random inclination angle $i$, obliquity angle $\beta$, and rotational phase,
as well as a white Gaussian noise with a null average and a variance
corresponding to the S/N of each observed profile. Using the fitted LSD $I$
profiles, we calculated local Stokes $V$ profiles assuming the weak-field case,
and we integrated over the visible hemisphere of the star. We obtained synthetic
Stokes $V$ profiles, which we normalized to the intensity of the continuum.
These synthetic profiles have the same mean Land\'e factor and wavelength as the
observations.

\begin{figure}[!ht]
\begin{center}
\resizebox{\hsize}{!}{\includegraphics[trim=1cm 1.5cm 4cm 2cm,clip]{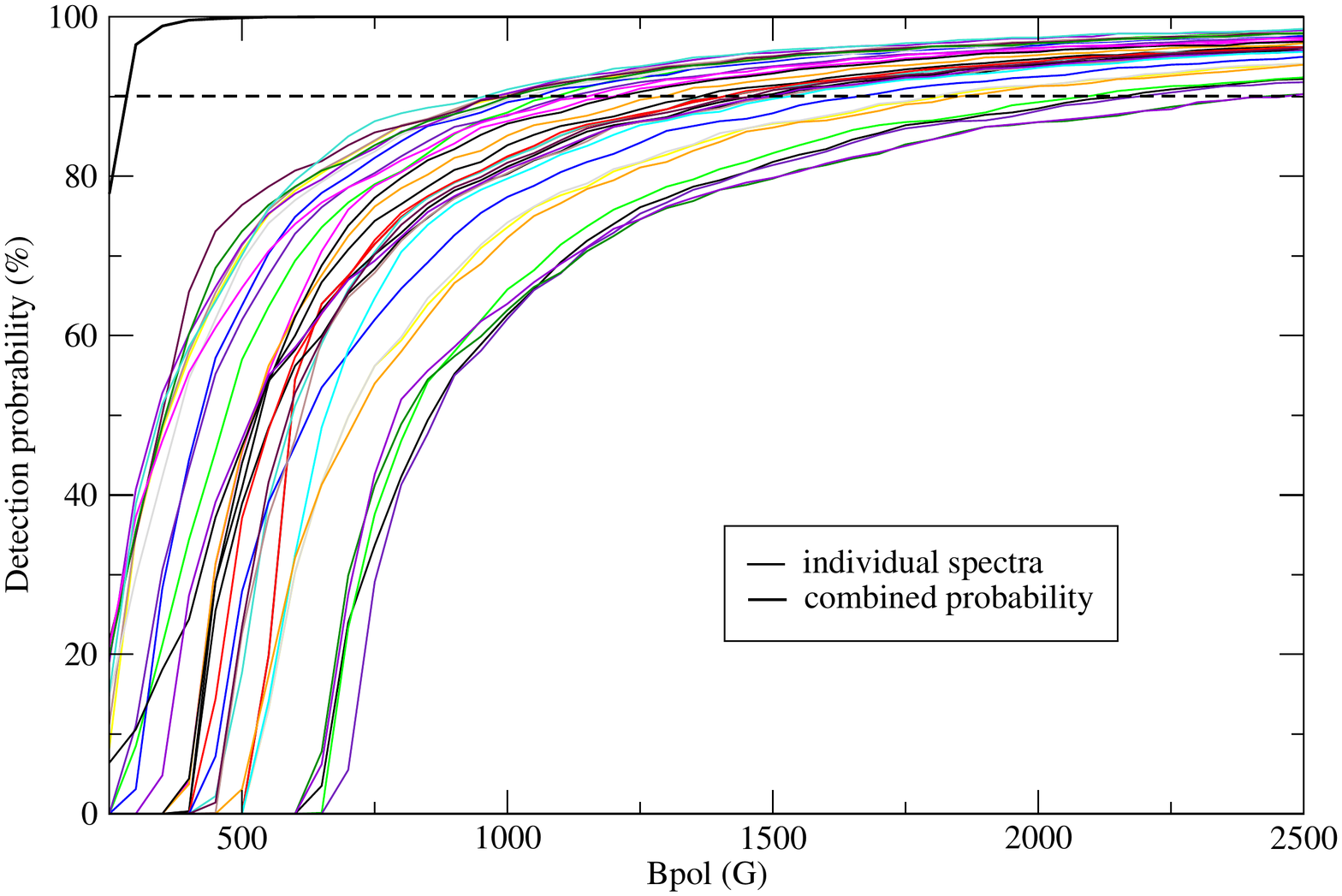}}
\caption[]{Detection probability of a magnetic field in each spectrum of the secondary component of $\zeta$\,Ori\,Ab (thin color lines) as a function of the magnetic polar field strength. The horizontal dashed line indicates the 90\% detection probability, and the thick black curve (top left corner) shows the combined probability.}
\label{limit}
\end{center}
\end{figure}

We then computed the probability of detecting a dipolar oblique magnetic
field in this set of models by applying the Neyman-Pearson likelihood ratio test
\citep[see e.g.][]{helstrom95,kay98,levy08}. This allowed us to decide between
two hypotheses: the profile only contains noise, or it contains a noisy Stokes
$V$ signal. This rule selects the hypothesis that maximizes the detection
probability while ensuring that the FAP is not higher than $10^{-3}$ for a
marginal magnetic detection. We then calculated the rate of detections in the
1000 models for each of the profiles depending on the field strength (see
Fig.~\ref{limit}).

We required a 90\% detection rate to consider that the field would statistically
be detected. This translates into an upper limit for the possible undetected
dipolar field strength for each spectrum, which varies between $\sim$900 and
$\sim$2350 G (see Fig.~\ref{limit}).

Since 36 spectra are at our disposal, statistics can be combined to extract a
stricter upper limit taking into account that the field has not been detected in
any of the 36  observations \citep[see][]{neiner15}. The final upper limit
derived from this combined probability for $\zeta$\,Ori\,Ab for a 90\% detection
probability is $\sim$300 G (see thick line in Fig.~\ref{limit}).

\section{Magnetic field configuration}\label{sect_config}
\subsection{Rotational modulation}\label{rot}

We searched for a period of variation in the 36 longitudinal magnetic field
measurements of $\zeta$\,Ori\,Aa with the clean-NG algorithm
\citep[see][]{gutierrezsoto09}. We obtained a frequency $f=0.146421$ c~d$^{-1}$,
which corresponds to a period of 6.829621 days. This value is consistent with 
the period of $\sim$7 days suggested by \cite{bouret08}. Assuming that the
magnetic field is a dipole with its axis inclined to the rotation axis, as is found
in the vast majority of massive stars, this period corresponds to the rotation
period of the star.

\begin{figure}
\centering

\includegraphics[scale=0.35,trim=0cm 2.98cm 0cm 2cm,clip]{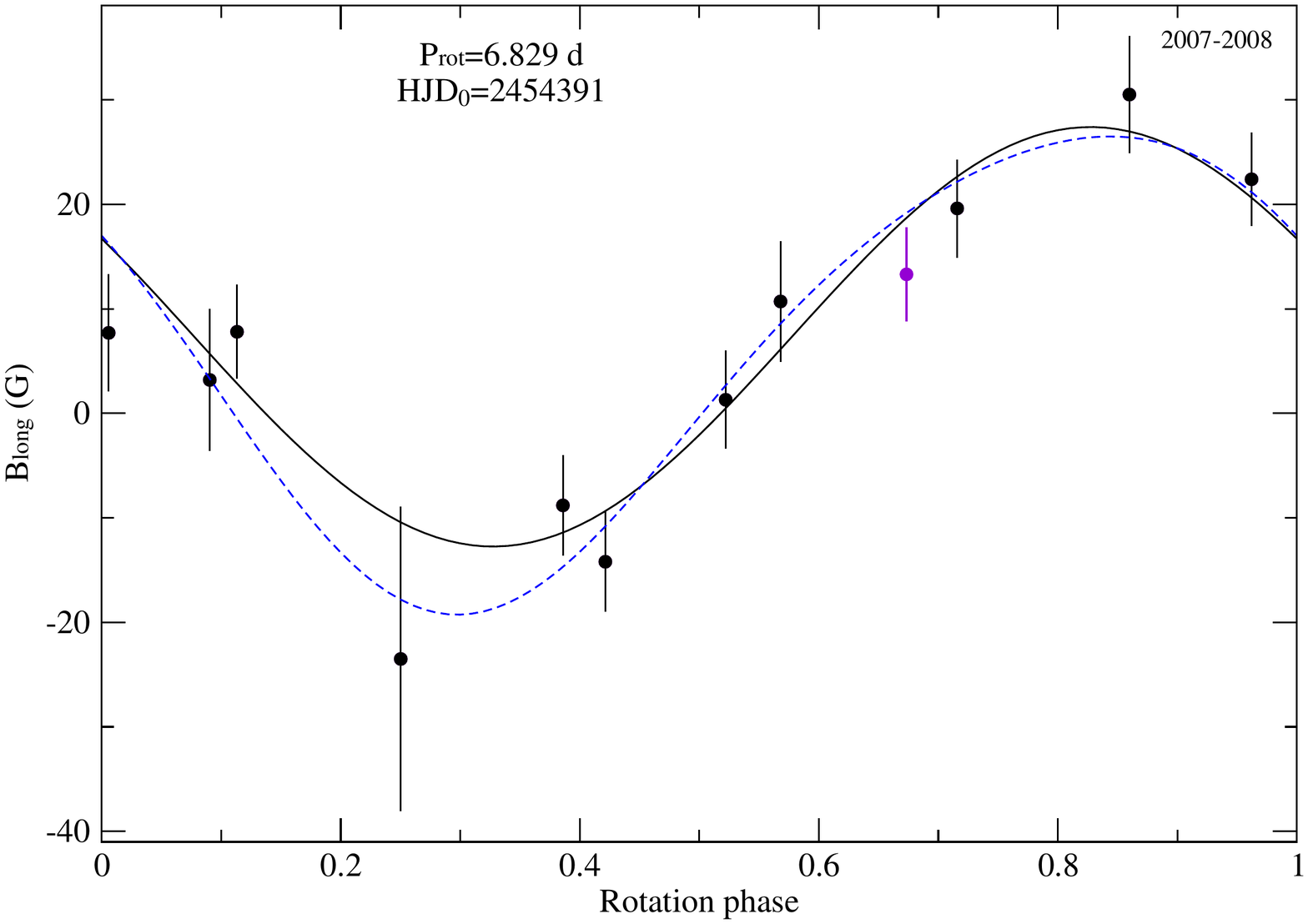}

\includegraphics[scale=0.35,trim=0cm 3cm 0cm 3.2cm,clip]{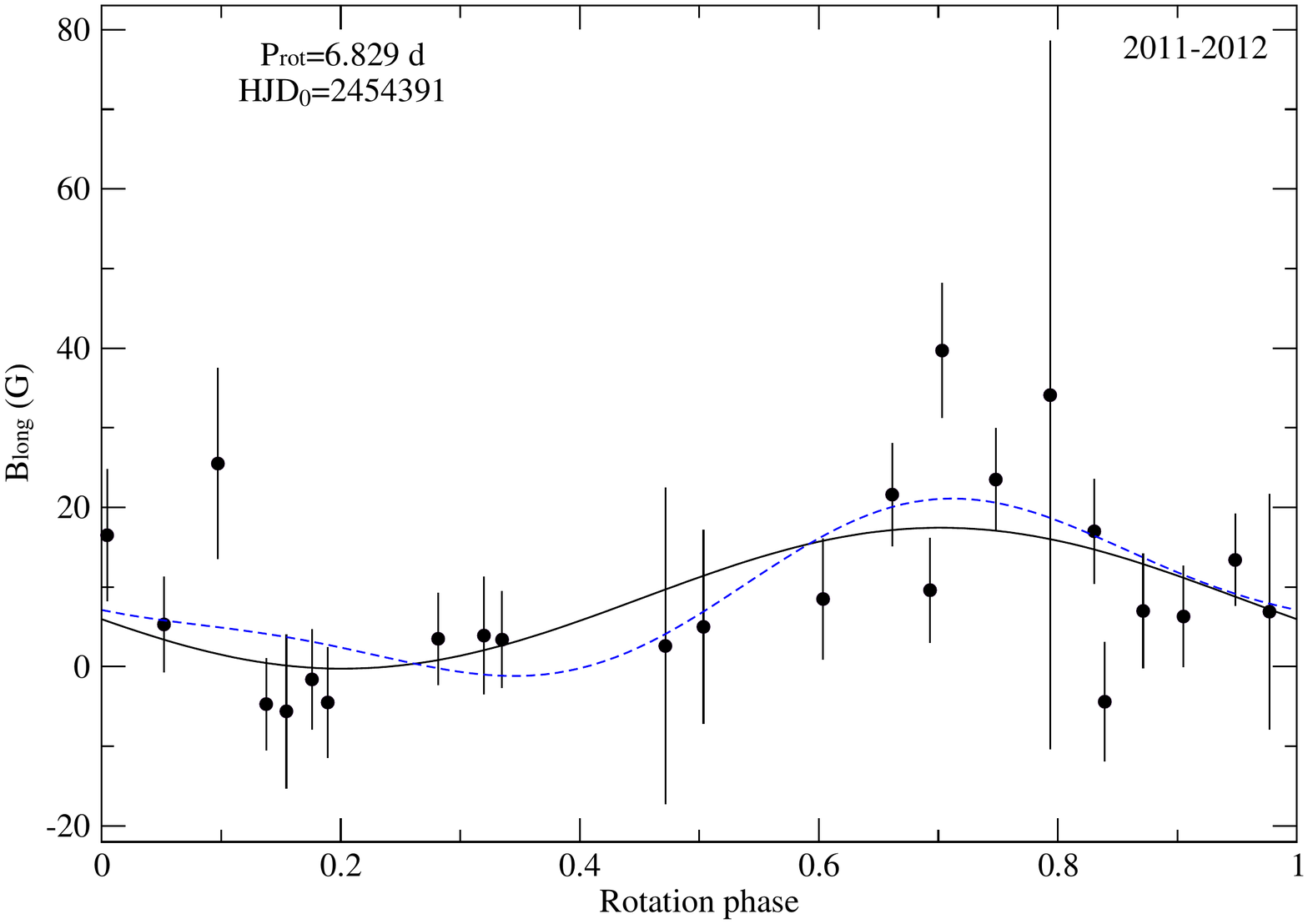}

\includegraphics[scale=0.35,trim=0cm 1cm 0cm 3cm,clip]{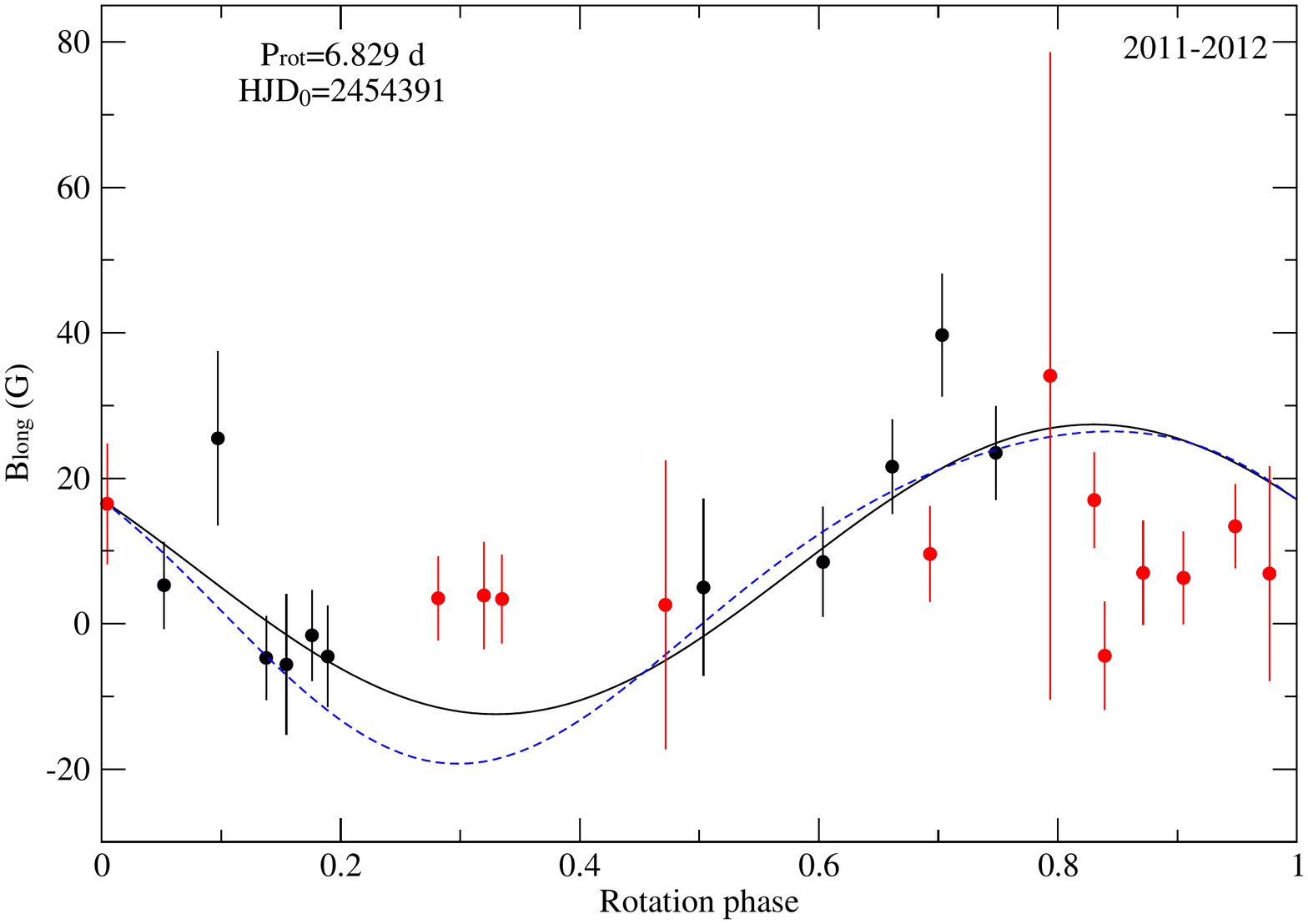}
\caption{Rotational modulation of the longitudinal magnetic field of $\zeta$\,Ori\,Aa for
the observations taken in 2007-2008 (top) and 2011-2012 (center). The black line
corresponds to the best dipolar fit, while the dashed blue line corresponds to
the best quadrupolar fit. The bottom panel compares the fit of the dipole and the quadrupole obtained from the 2007-2008 data with the observations obtained in 2011-2012 (see Sect.~\ref{vmodel}). The data for which the Stokes V model matches the observed LSD V profiles are shown in black, while the data for which the Stokes V model does not match are in red.}
\label{period}
\end{figure}

We used this period and plotted the longitudinal magnetic field as a function of
phase. For the data taken in 2007 and 2008, the phase-folded field measurements
show a clear sinusoidal behavior, as expected from a dipolar field model (see
top panel of Fig.~\ref{period}). A dipolar fit to the data, that is, a sine fit of
the form $B(x) = B_0 + B_a \times sin(2\pi(x+\phi_d))$, resulted in $B_0$=6.9 G
and $B_a$=19.2 G. A quadrupolar fit to the phase-folded data only shows
an insignificant departure from the dipolar fit.

However, the period of $\sim$6.829 days does not to match the measurements
collected in 2011 and 2012 very well (see middle panel of Fig.~\ref{period}).
None of the dipolar or quadrupolar fits to these data provide a reasonable
match. A further search for a different period in these 2011-2012 data alone provided no significant result.

The magnetic fields of main-sequence massive stars are of fossil origin. These fields are
known to be stable over decades and are only modulated by the rotation of the star.
A change of period in the field modulation between the 2007-8 and 2011-12 epochs
is thus not expected in $\zeta$\,Ori\,Aa. 

$\zeta$\,Ori\,Aa has a companion, therefore we investigated the
possibility that the magnetic field has been affected by the companion. Indeed,
in 2011 and 2012, $\zeta$\,Ori\,Ab was close to periastron, which means that the distance
between the two stars was smaller than in 2007 and 2008. We calculated this
distance to check whether some binary interactions might have occurred.

To calculate the distance between the two components, we used the photometric
distance of $\zeta$\,Ori\,A, d=387 pc \citep{hummel13}. From
\cite{hummel13}, we know the orbital parameters of the binary. The shortest
distance between the two stars is $r_{\rm min}= a-\sqrt{a^2-b^2}$= 23.8 mas, where
a is the semi-major axis and b the semi-minor axis. From the distance of
$\zeta$\,Ori\,A, we can compute $r_{\rm min}=\sin(\theta)/d$ in pc, where
$\theta$ is the parallax in radian. We obtained a distance of 81 R$_{*}$, where
R$_{*}$ is the radius of $\zeta$\,Ori\,Aa.

The distance between the two stars at periastron therefore appears too large for
interactions between the two stars to occur. In addition, the binary system is
still significantly eccentric \citep[0.338,][]{hummel13} even though $\zeta$\,Ori\,Aa has already evolved into a supergiant. Tidal interactions
have apparently not been able to circularize the system yet, which would confirm
that these interactions are weak \citep{zahn08}.

However, in addition to $\zeta$\,Ori\,Aa and Ab, a third star $\zeta$\,Ori\,B
may also interfere with the $\zeta$\,Ori\,A system. \cite{correia12} showed
that when a third component comes into play, tidal effects combined with
gravitational interactions may increase the eccentricity of $\zeta$\,Ori\,A,
which would otherwise have circularized. We thus cannot exclude that tidal
interactions are stronger than they seem in the $\zeta$\,Ori\,A system.

\subsection{Field strength and geometrical configuration}\label{geom}

Assuming that the period detected in Sect.~\ref{rot} is the rotation period of
the star, we can determine the inclination angle of the star $i$ by measuring
\vsini. In massive stars, line broadening does not come from rotational
broadening alone, but also from turbulence and stellar wind. This is particularly
true for supergiant stars. 

Based on the synthetic spectra calculated in Sect.~\ref{lines} and the lines
identified to belong to only one of the two components, we determined  the
broadening needed in the synthetic spectra to fit the observations. For
$\zeta$\,Ori\,Aa, a broadening of 230 \kms\ was necessary to provide a good fit
to the observed lines, while for $\zeta$\,Ori\,Ab we needed 100 \kms. These
broadening values are upper limits of the \vsini\ values because they include all physical processes that broaden the lines. In fact, with a period
of 6.829 days and a radius of 20 R$_\odot$ as given by \cite{hummel13}, the
maximum possible \vsini\ for $\zeta$\,Ori\,Aa is 148 \kms. 

In addition, \cite{bouret08} determined \vsini\ through a Fourier transform of
the average of the 5801 and 5812 \AA\ CIV and 5592 \AA\ OIII line profiles. They
found a \vsini\ of 110$\pm$10 \kms. From our disentangling of the spectra,
we know that the two CIV lines originate from $\zeta$\,Ori\,Aa, but the OIII
5590 \AA\ line is partly ($\sim$10\%) polluted by $\zeta$\,Ori\,Ab. As a
consequence, we applied the Fourier transform method to the LSD Stokes I
profiles we calculated from the lines that only originate from $\zeta$\,Ori\,Aa. We
obtained \vsini = 140 \kms. However, it is known that \vsini\ values determined
from LSD profiles might be overestimated.

Finally, taking macroturbulence into account but not binarity, foe example,
\cite{simondiaz14} found that \vsini\ for $\zeta$\,Ori\,A is between
102 and 127 \kms, depending on the method they used. 

In the following, we thus consider that \vsini\ is between 100 \kms and 148 \kms
for $\zeta$\,Ori\,Aa. In addition, we adopt the radius of 20 R$_\odot$ given by
\cite{hummel13} and the rotation period of 6.829 d. Using \vsini = [100-148]
\kms, we obtain $i \sim [42-87]^\circ$.

Using the dipolar fit to the 2007-2008 longitudinal field measurements and the
inclination angle $i$, we can deduce the obliquity angle $\beta$ of the magnetic
field with respect to the rotation axis. To this aim, we  used the formula $r =
B_{\rm min} / B_{\rm max} = \cos(i - \beta) / \cos(i + \beta)$
\citep{shore87}. The dipolar fit of the longitudinal field values gives $r$=0.47. With $i \sim [42-87]^\circ$,
we obtain $\beta \sim [71-8]^\circ$. 

In addition, from the dipolar fit to the longitudinal field values and the
angles $i$ and $\beta$ determined above, we can estimate the polar field
strength with the formula $B_0 \pm B_a = 0.296 \times B_{\rm pol} \cos(\beta \pm
i)$, where the limb-darkening coefficient is assumed to be 0.4
\citep[see][]{borra80}. We found $B_{\rm pol}$ = $[110\pm5-524\pm65]$ G. The
dipolar magnetic field that we find is thus higher than the one found by
\cite{bouret08}.

In 2011-2012, the maximum measured B$_l$ is 51 G and the minimum polar field
strength is thus $B_{\rm pol} \geq 3.3 B_{l,max} = 168\pm33$ G. This value is
compatible with the range derived from the 2007-2008 data.

\subsection{Stokes V modeling}\label{vmodel}

Since the $B_l$ data taken in 2007-2008 point toward the presence of a dipole
field, we used an oblique rotator model to fit the LSD Stokes V and I profiles.

We used Gaussian local intensity profiles with a width calculated according to
the resolving power of Narval and a macroturbulence value of 100 \kms\
determined by \cite{bouret08}. We fit the observed LSD I profiles by Gaussian
profiles to determine the depth, \vsini\  and radial velocity of the intensity
profile. We used the weighted mean Land\'e factor and wavelength derived from
the LSD mask applied to the Narval observations and the rotation period of 6.829
days. The fit includes five parameters: i, $\beta$, B$_{\rm pol}$, a phase shift
$\phi$, and a possible off-centering distance d$_d$ of the dipole with respect  to
the center of the star (d$_d$=0 for a centered dipole and d$_d$=1 if the center of the
dipole is at the surface of the star).

\begin{figure*}
\centering

\includegraphics[scale=0.8,trim=0.0cm 0cm 0cm 0cm,clip]{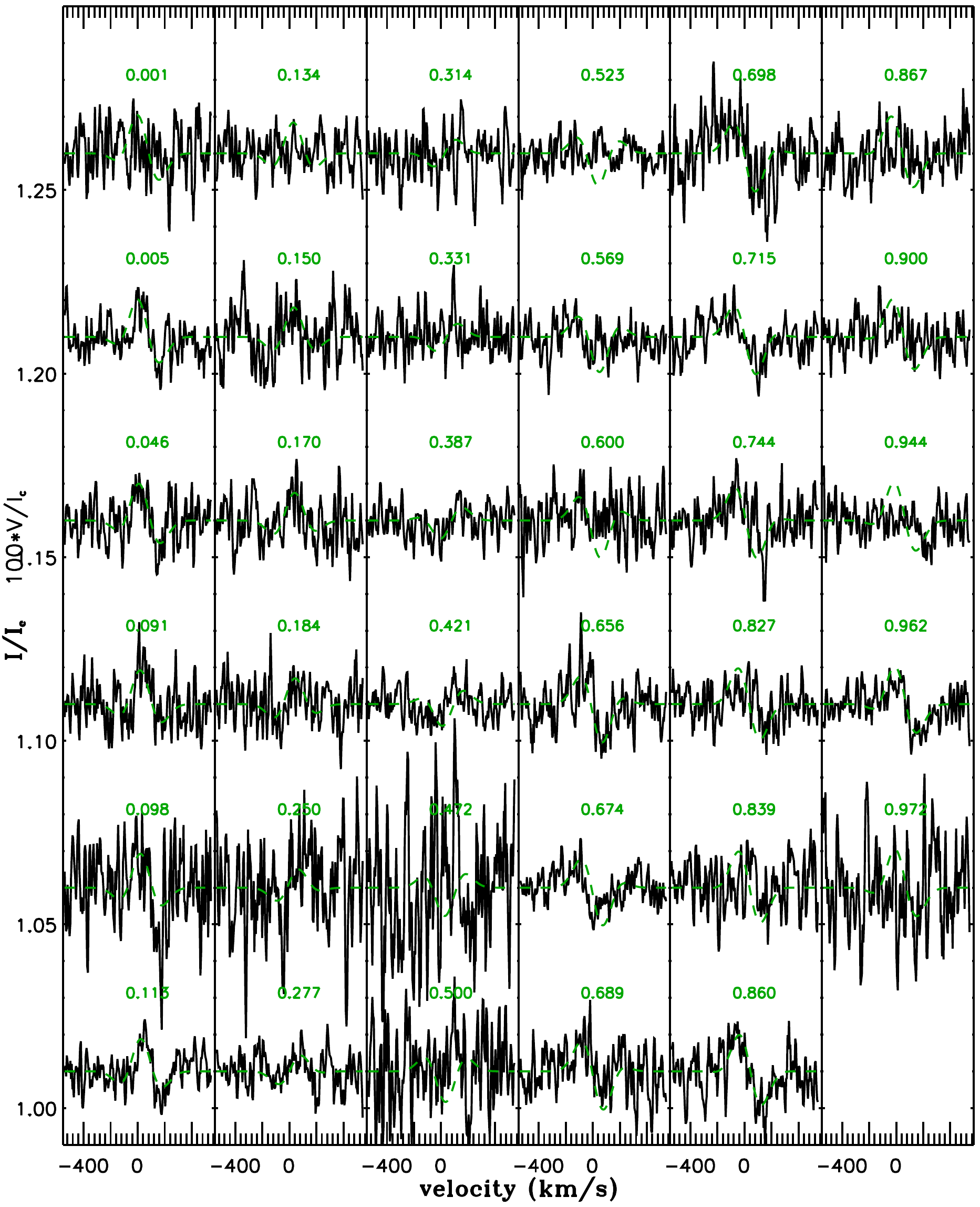}
\caption{Best dipolar model fit (green) of the observed Stokes V profiles (black). The green numbers correspond to the rotational phase. The very noisy observation at phase 0.793 is not shown.}
\label{model_V}
\end{figure*}

\section{Magnetospheres}
\label{sect_magnetosphere}
\subsection{Magnetospheric parameters}
We calculated a grid of V profiles for each phase of observation by varying the
five parameters mentioned above and applied a $\chi^2$ minimization to obtain
the best fit of all observations simultaneously. More details of the modeling
technique can be found in \cite{alecian08}. The parameters of the best fit are
i=79.89$^\circ$, $\beta$=21.5$^\circ$, $\phi$=0.68, B$_{\rm pol}$=142.2 G and
d$_d$=0.0. The values for the angles $i$ and $\beta$ are within the error boxes
derived in Sect.~\ref{geom}, and the value for the polar field strength B$_{\rm
pol}$ fits the $B_l$ results well. Moreover, the best fit is obtained for
d$_d$=0, which confirms that no quadrupolar component is found.

The 36 Stokes V profiles for this best fit are shown in Fig.~\ref{model_V} overplotted on the observations. As expected, for the nights in 2007-2008, the model
fits the observations well. For some nights in 2011-2012, the observations are
too noisy to see whether the model fits well. Considering the nights in
2011-2012 for which the S/N is sufficient, the model fits some the observations
but not all. For those nights for  which the model fitted well the observations,
we compared the values of the longitudinal magnetic field B$_l$ to the dipolar
fit obtained for the B$_l$ measurements of 2007-2008 (see bottom panel of
Fig.~\ref{period}). The 2011-2012 data that match the Stokes V models also match
the $B_l$ dipolar fit curve. Therefore, it seems that at least part of the
2011-2012 data show the same rotational modulation and dipole field as in
2007-2008. Only part of the 2011-2012 dataset does not match the rest of the
observations.

With the polar magnetic field strength $B_{\rm pol}$ = 142.2 G determined with
to the Stokes V model, we calculated the wind confinement parameter $\eta_*$
which characterizes the ability of the magnetic field to confine the wind
particles into a magnetosphere \citep{uddoula02}. If $\eta_* \leq 1$,
$\zeta$\,Ori\,Aa is located in the weakly magnetized winds region of the
magnetic confinement-rotation diagram and it does not have a magnetosphere.
However, for $\eta_* > 1$, wind material is channeled along magnetic field lines
toward the magnetic equator and $\zeta$\,Ori\,Aa hosts a magnetosphere.

To calculate $\eta_*$, we first used the fiducial mass-loss rate
$\dot{M}_{B=0} =1.4 \times 10^{-6} M_\odot~yr^{-1}$ and the terminal speed V$_\infty = 2100$
\kms\ determined by \cite{bouret08}. They measured the mass-loss rate from
the emission of H$\alpha$ and used archival International Ultraviolet
Explorer (IUE) spectra to measure the wind terminal velocity from the blueward
extension of the strong UV P Cygni profile. We obtain $\eta_*$ = 0.9. 

We then recalculated $\eta_*$ but this time using the mass-loss rate of
$\dot{M}=3.4 \times 10^{-7} M_\odot~yr^{-1}$ and V$_\infty = 1850$ \kms\ determined by
\cite{cohen14}. This gives $\eta_*$ = 4.2.

A magnetosphere can only exist below the Alfv\'en radius $R_A$, which is
proportional to $\eta_*$, with $R_A = \eta_*^{1/4} R_*$. For $\zeta$\,Ori\,Aa,
using the two above determinations of $\eta_*$, $R_A = [0.98-1.43] R_*$.
Moreover, the magnetosphere can be centrifugally supported above the corotation
Kepler radius $R_K$. $R_K = (2\pi R_* / P_{\rm rot} \sqrt(GM/R_*) )^{2/3}$, thus
for $\zeta$\,Ori\,Aa $R_K = 2.8 R_*$. Since $R_K$ > $R_A$, no centrifugally
supported magnetosphere can exist.

Therefore, $\zeta$\,Ori\,Aa is either in the weakly magnetized winds region of
the magnetic confinement-rotation diagram, meaning that $\zeta$\,Ori\,Aa does not have a
magnetosphere ($\eta_* < 1$), or it hosts a dynamical magnetosphere ($1 < \eta_*
< 4.2$). 

\subsection{H$_\alpha$ variations}

The H$_\alpha$ line shows significant variability in emission and absorption.
For stars that have a magnetosphere, we expect magnetospheric emission at
H$_\alpha$, which varies with the rotation period \citep[see
e.g.][]{grunhut15}. 

\begin{figure}
\centering
\includegraphics[scale=0.3,trim=0.5cm 0.07cm 0cm 0cm,clip]{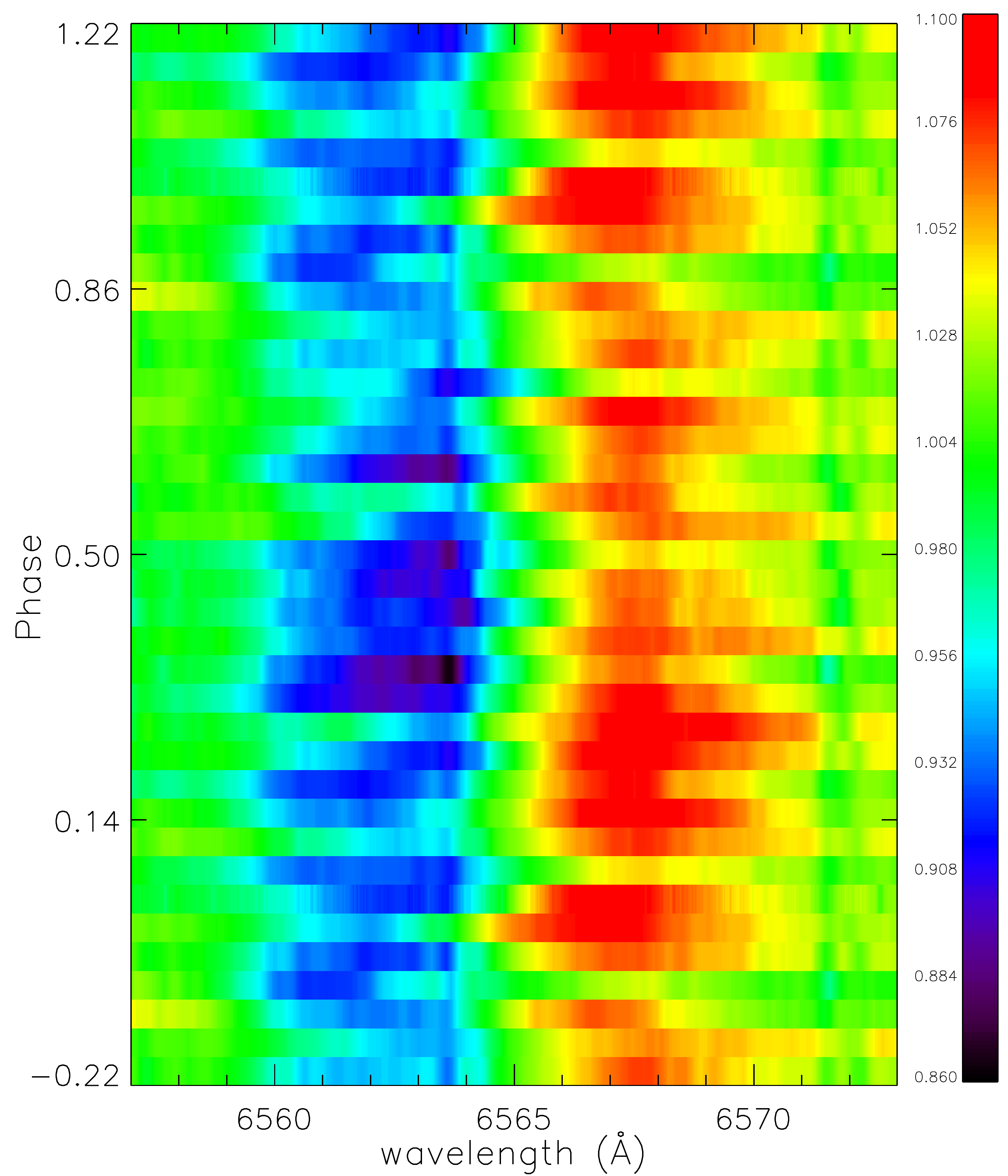}
\caption{Dynamic plot of each individual archival H$_\alpha$ spectrum in phase with the rotation period.}
\label{h_alpha}
\end{figure}

\begin{figure*}
\centering
\includegraphics[scale=0.6,trim=0cm 0.0cm 0cm 0cm,clip]{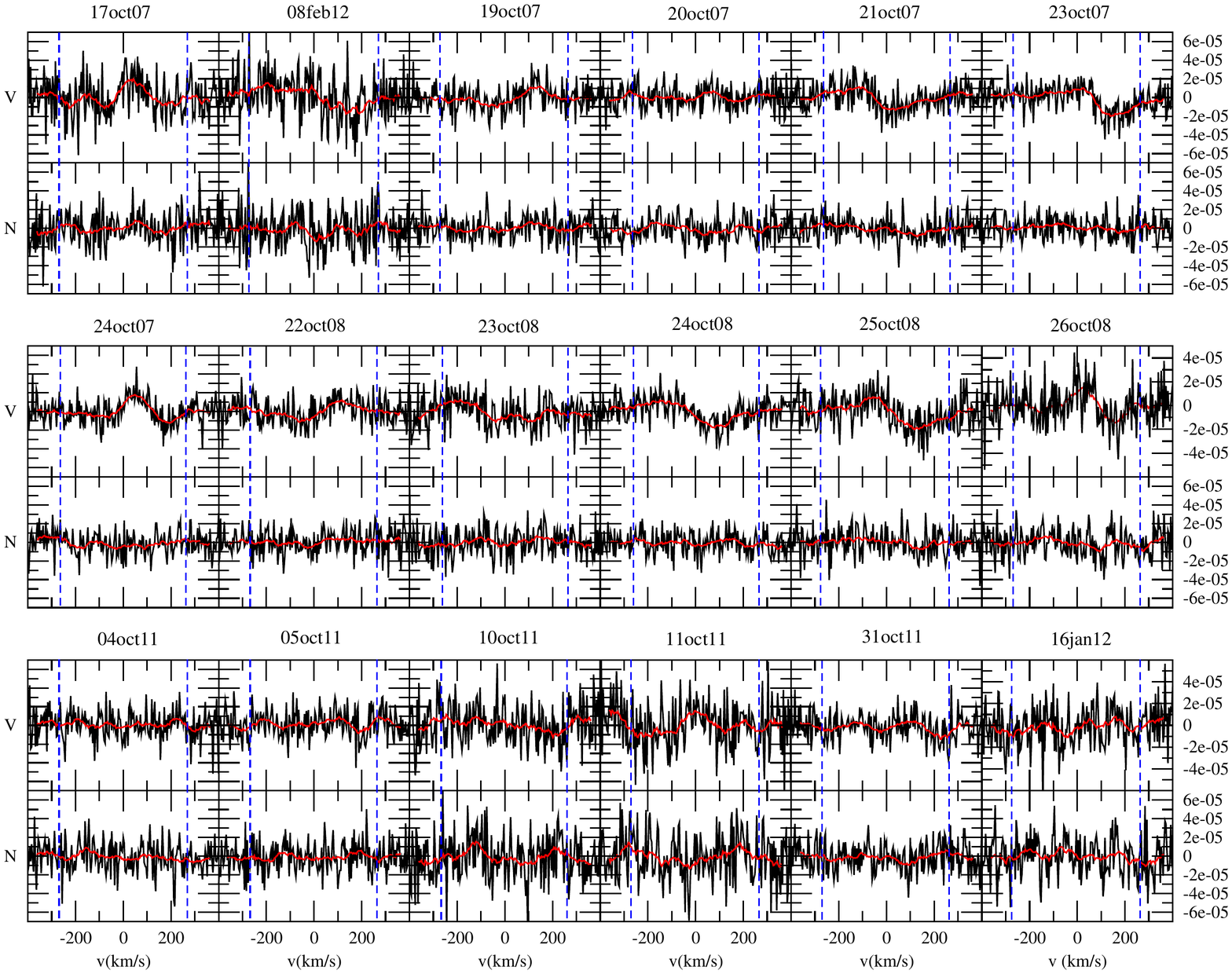}
\includegraphics[scale=0.6,trim=0cm 0.5cm 0cm 0cm,clip]{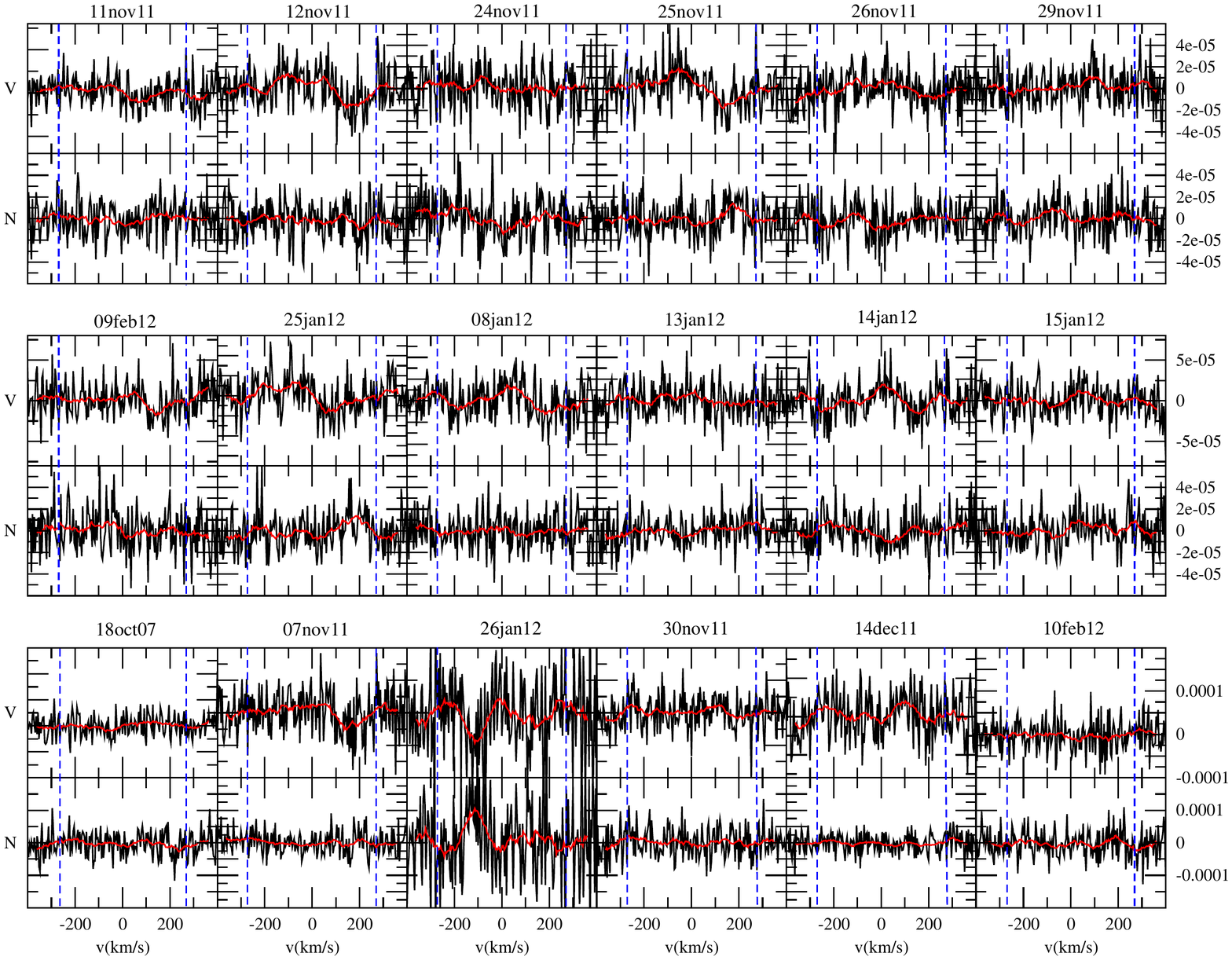}
\caption{LSD Stokes I profiles (bottom) computed from the disentangled spectroscopic data, Stokes V (top) and null N (middle) profiles, normalized to Ic, from the Narval data, for $\zeta$\,Ori\,A. The red line is a smoothed profile.}
\label{lsd3}
\end{figure*}

To check whether there is a signature of the presence of a magnetosphere around
$\zeta$\,Ori\,Aa, we studied the variation of its H$_\alpha$ line in the archival spectra (see Sect.~\ref{sect_spectro}). We confirm
that the emission in H$_\alpha$ does indeed vary. While most of the variations are problably related to variations in the stellar wind of the supergiant, the signature of a weak rotationally modulated dynamical magnetosphere is observed in H$_\alpha$ (see Fig.~\ref{h_alpha}).

The ratio $\log R_A/R_K$ gives a measure of the volume of the magnetosphere. For $\zeta$\,Ori\,Aa, $\log R_A/R_K$ is very small ($<$ 0.06), and it is thus not surprising that H$_\alpha$ only  weakly reflects magnetic confinement.

\section{Discussion and conclusions}\label{sect_discussion}

Based on archival spectrocopic data and Narval spectropolarimetric data, we
confirm the presence of a magnetic field in the massive star $\zeta$\,Ori\,A, as
initally suggested by \cite{bouret08}. However, \cite{bouret08} ignored that
$\zeta$\,Ori\,A is a binary star, which was subsequently shown by \cite{hummel13} with  
interferometry.

We disentangle the spectra and could thus show that the primary O 
supergiant component $\zeta$\,Ori\,Aa is the magnetic star, while the secondary
$\zeta$\,Ori\,Ab is not magnetic at the achieved detection level. 
$\zeta$\,Ori\,Aa is the only magnetic O supergiant known as of today. 

The magnetic field of $\zeta$\,Ori\,Aa is a typical oblique dipole field,
similar to those observed in main-sequence massive stars. From Stokes modeling,
the polar magnetic field strength $B_{\rm pol}$ of $\zeta$\,Ori\,Aa is found to
be about 140 G. If we assume field conservation during the evolution of $\zeta$\,Ori\,Aa because the stellar radius increased from $\sim$10 to $\sim$20 R$_\odot$,
the surface magnetic polar field strength decreased by a factor $\sim$4. This
implies that the polar field strength of $\zeta$\,Ori\,Aa when it was on the
main sequence was about 600 G. This is similar to what is observed in other main-sequence magnetic O stars.

The current field strength and rotation rate of $\zeta$\,Ori\,Aa are weak, with respect to the wind energy, for the star to be able to host a centrifugally
supported magnetosphere. However, it seems to host a dynamical magnetosphere. All other
ten known magnetic O stars host dynamical magnetospheres, except for the complicated
system of Plaskett's star, which has a very strong magnetic field and hosts a
centrifugally supported magnetosphere \citep[see][]{grunhut13}. However, these
other magnetic O stars are not supergiants.

Although $\zeta$\,Ori\,A is one of the brightest O star in the X-ray domain, 
\cite{cohen14} found that it resembles a non-magnetic star, with no evidence
for magnetic activity in the X-ray domain and a spherical wind. 
This probably results from the weakness of the magnetosphere around $\zeta$\,Ori\,Aa.

The rotation period of $\zeta$\,Ori\,Aa, $P_{\rm rot}$ = 6.829 d, was determined
from the variations of the longitudinal magnetic field. This period is clearly
seen in the data obtained in 2007-2008, but only part of the 
spectropolarimetric measurements obtained in 2011 and 2012 seem to follow that
rotational modulation. The reason for the lack of periodicity for part of the
magnetic measurements of 2011-2012 was not identified. Although passage at the
binary periastron occured between 2008 and 2011, the distance between the two
companions seems too large for the companion to have perturbed the magnetic
field of the primary star, unless it is $\zeta$\,Ori\,B which has maintained the
two components of $\zeta$\,Ori\,A at a distance (see Sect.~\ref{rot}).

$\zeta$\,Ori\,A therefore remains an interesting star that needs to be studied
further. More spectropolarimetric observations should be collected at
appropriate orbital phases to allow for a more accurate spectral disentangling.
this would allow obtaining stronger constraints on the magnetic field strength and
configuration, studying the field as a function of orbital phase, and understanding
the magnetic field perturbations that seem to have occurred during the
observations in 2011-2012.

\begin{acknowledgements}
AB thanks Patricia Lampens and Yves Fr\'emat for useful discussions on the
disentangling technique. AB and CN also thank St\'ephane Mathis for valuable 
discussions on tidal effects and Fabrice Martins for helpful discussion about the spectral classification. AB and CN acknowledge support from the Agence
Nationale de la Recherche (ANR) project Imagine. This research has made use of the
SIMBAD database operated at CDS, Strasbourg (France), and of NASA's Astrophysics
Data System (ADS).
\end{acknowledgements}

\bibliographystyle{aa} 
\bibliography{biblio}

\Online

\end{document}